\documentclass[amsmath,amssymb,nofootinbib,superscriptaddress]{revtex4-1}

\usepackage{esint,graphicx}

\usepackage{soul}
\usepackage{comment}
\usepackage[compact]{titlesec}
\titlespacing{\subsection}{10pt}{10pt}{10pt}
\titlespacing{\section}{10pt}{10pt}{10pt}
\usepackage[allcolors=blue,colorlinks=true]{hyperref}
\usepackage[T1]{fontenc}
\usepackage[textsize=footnotesize,textwidth=2.cm]{todonotes}
\usepackage{float}
\usepackage{braket}

\numberwithin{equation}{section}

\makeatletter
\renewcommand{\p@subsection}{}
\renewcommand{\p@subsubsection}{}
\makeatother

\newcommand{\p}{\mathop{}\!\partial}

\newcommand{\ba}{\begin{array}}
\newcommand{\ea}{\end{array}}
\newcommand{\bi}{\begin{itemize}}
\newcommand{\ei}{\end{itemize}}
\newcommand{\bea}{\begin{eqnarray}}
\newcommand{\eea}{\end{eqnarray}}
\newcommand{\be}{\begin{equation}}
\newcommand{\ee}{\end{equation}}
\newcommand{\nn}{\nonumber}

\begin{document}

\begin{flushright}
USTC-ICTS/PCFT-25-44
\end{flushright}

\title{\vspace*{40pt}\Large{Logarithmic Corrections to Thermodynamics of Accelerating Black Holes}\vspace*{40pt}}

\author{Jianfei Xu}
\email{jfxu@seu.edu.cn}
\affiliation{Shing-Tung Yau Center and School of Mathematics, Southeast University, Nanjing, 210096, China}
\affiliation{Peng Huanwu Center for Fundamental Theory, Hefei, Anhui, 230026, China}

\begin{abstract}
\vspace*{40pt}
As pointed out in recent research, the near extremal black hole entropy with one-loop effect exhibits universal $\log T$ behavior at sufficiently low temperature. In this paper, we discuss the low-temperature quantum corrections to the thermodynamics of four-dimensional accelerating black holes with rotation and charges by using the method of Euclidean path integral. The one-loop path integral for the black hole thermal partition function near extremality is dominated by zero modes defined with respect to the extremal background. For the accelerating black holes without rotation, the near horizon extremal geometry is a direct product of AdS$_2$ and S$^2$ with warping factors, and the gravitational zero modes contain both tensor and vector types, with the respective contributions to the near extremal black hole entropy being $(3/2)\log T$ and $(1/2)\log T$. While in the presence of rotation, the near horizon extremal geometry is a warped and twisted product of AdS$_2$ and S$^2$ and the gravitational vector modes are absent. For the accelerating black holes with charges, we also consider the one-loop path integral of the gauge field, where the photon zero modes are found to contribute an additional $(1/2)\log T$ term to the near extremal black hole entropy.
\end{abstract}

\maketitle

\clearpage

{
  \hypersetup{linkcolor=black}
  \tableofcontents
}

\vspace{30pt}

\section{Introduction}
Black holes are fascinating thermal objects with their own laws of thermodynamics that are analogous to those of classical thermodynamics. From the semiclassical analysis, the temperature is determined by the surface gravity and the entropy is proportional to the area of the horizon of the black hole~\cite{Bekenstein:1973ur, Hawking:1976de, Gibbons:1976ue, York:1986it}. However, black holes are distinct from ordinary thermal systems. As two of the Killing horizons converge, the temperature of a charged or rotating black hole can become very small, while the entropy remains finite value since the black hole horizon has nonvanishing area in the extremal limit. This phenomenon indicates there exists a large amount of ground-state degeneracy with the density of black hole states proportional to the exponential of its horizon area. For supersymmetric black holes in string theory, the thermal entropy in the extremal cases are well-understood by the microscopic state counting~\cite{Strominger:1996sh, Sen:2014aja}. For nonsupersymmetric black holes like those in our Universe, such a large degeneracy need further explanations. Whether there exists an unknown symmetry protecting mechanism at low temperature to ensure the ground state degeneracy or it is simply a result of the semiclassical approximation remains an open question for thorough investigation.

One step forward is to incorporate quantum corrections into the calculation of the black hole entropy when going beyond the semiclassical approximation in the gravitational path integral. For extremal black holes, the leading quantum correction to the entropy is found to take a universal form of logarithm of the horizon area~\cite{Banerjee:2010qc, Banerjee:2011jp, Sen:2012kpz, Sen:2012cj}, without invoking the explicit ultraviolet structure of the underlying quantum theory of gravity. In the nonextremal cases, a complete profile of the finite temperature quantum correction still hide in mist. Once the temperature is very small, the near horizon geometry of any near extremal black hole in four dimensions universally takes the form as a product of two dimensional anti-de Sitter space (AdS$_2$) and two sphere (S$^2$). With this property, the quantum fluctuations above the extremal or near extremal configurations exhibit controllable behaviors. As pointed out in~\cite{Iliesiu:2020qvm, Heydeman:2020hhw, Boruch:2022tno, Iliesiu:2022onk, Rakic:2023vhv}, the quantum fluctuations in the near horizon region of higher-dimensional extremal or near extremal black holes are well captured by the two-dimensional Jackiw-Teitelboim (JT) gravity~\cite{Jackiw:1984je, Teitelboim:1983ux} through dimensional reduction and the low energy physics is described by a Schwarzian theory~\cite{Almheiri:2014cka, Maldacena:2016upp} with certain boundary conditions of the AdS$_2$ factor in the near horizon geometry. By using the path integral of the Schwarzian theory at low temperature, $\log T$ corrections are found in the thermal entropy of the theory, which is different from the logarithm of the horizon area correction to the extremal entropy. The $\log T$ term in entropy is essential to resolve the mass gap puzzle~\cite{Preskill:1991tb} appears in the low-temperature region of nonextremal black holes under the semiclassical approximation. Without quantum corrections, the black hole thermal energy above extremality scales quadratically in small temperature while the average energy of the Hawking radiation scales linearly in temperature. This means that the black hole can not radiate even a single Hawking quantum at sufficiently low temperature. The $\log T$ quantum correction to the entropy in turn corrects the thermal energy above extremality with a term linear in small temperature, and thus address the issue involved in the mass gap puzzle.

The $\log T$ corrections can also be figured out by using the Euclidean path integral without resorting to the dimensional reduction. In Ref.~\cite{Banerjee:2023quv}, the authors evaluate the one-loop partition function for the near extremal black hole in the Einstein-Maxwell theory by using the Euclidean path integral. A set of zero modes with zero eigenvalues of the kinetic operator of small fluctuations are found to contribute leading quantum corrections in the path integral. In their analysis, the near extremal black holes are viewed as small temperature deviations form the extremal ones. The small temperature perturbation regularize the divergent contribution from the zero modes path integral on the extremal background and result a $\log T$ term in the logarithm of the partition function. The same logic also applies to the Kerr black holes~\cite{Kapec:2023ruw}, where the near horizon extremal Kerr (NHEK) geometry plays a crucial role in determining the zero modes in the Euclidean path integral. Interestingly, these zero modes are generated by the vector fields which are asymptotic to those large diffeomorphisms in the Kerr/CFT correspondence~\cite{Guica:2008mu, Castro:2009jf, Bredberg:2011hp}. See~\cite{Maulik:2024dwq, Maulik:2025phe} for the discussions of the universality of the low-temperature quantum corrections for the asymptotically AdS and dS black holes by using the gravitational path integral method. For charged black holes with additional gauge fields, the quantum corrections from the gauge fields path integral should also be taken into account. It is found in~~\cite{Blacker:2025zca} that the gauge zero modes also contribute in the Euclidean path integral and lead to additional $\log T$ corrections to the thermal entropy. Thus, the coefficient of the $\log T$ term depends on the near horizon geometry of the extremal black hole as well as the field contents in the gravity theory.

In this paper, we will follow the spirit of~\cite{Kapec:2023ruw} and consider the low-temperature quantum corrections for accelerating black holes. An accelerating black hole is usually known as the C-metric which describes a pair of black holes constantly accelerating away from each other~\cite{Kinnersley:1970zw, Plebanski:1976gy, Dias:2002mi, Hong:2003gx, Hong:2004dm, Griffiths:2005se, Griffiths:2005qp, Podolsky:2006px}. Without specifying the range of the polar angle, the metric of an accelerating black hole contains deficit angles at the north and south poles. Therefore, its acceleration can be effectively viewed as the result of cosmic strings attached at the poles with tensions determined by the deficit angles. Much like the Kerr case, a rotating black hole with acceleration also has an infinite near horizon region at extremality. For example, the extremal accelerating Kerr-Newmann black hole is shown to acquire a warped and twisted product of AdS$_2$ and S$^2$ near horizon scaling region, where the asymptotic symmetry analysis can be applied to realize a 2D conformal field theory (CFT) dual~\cite{Astorino:2016xiy}. The thermodynamic properties of accelerating black holes have been widely explored~\cite{Appels:2016uha, Astorino:2016ybm, Appels:2017xoe, Anabalon:2018ydc, Zhang:2018hms, Anabalon:2018qfv, Gregory:2019dtq, Ball:2020vzo, Ball:2021xwt, Cassani:2021dwa, Kim:2023ncn, Wu:2023meo, Liu:2025iyl, Hale:2025veb}. We will use the Euclidean path integral to figure out the low-temperature quantum corrections to the thermodynamics of accelerating black holes, considering both rotating and charged cases. The existences of the AdS$_2$ factor in their near horizon extremal geometry entails that a similar statement regarding the low-temperature quantum corrections holds true in these cases. We will highlight the effect due to the acceleration in comparing with the nonaccelerating cases.

This paper is organized as follows. In Sec. \ref{sec2}, we first review the the thermodynamics of the accelerating Kerr black hole, and the low-temperature behaviors of the thermodynamic quantities. We carry out the near horizon geometry of the near extremal accelerating Kerr black hole with an infinite scaling coordinate transformation and solve the zero modes of the kinetic operator in the one-loop Euclidean path integral on the extremal background. Then we perturb the extremal background with the first-order temperature perturbation and calculate the finite temperature corrections to the eigenvalue of the zero modes. These finite and nonvanishing eigenvalues determine the leading one-loop quantum corrections to the logarithm of the partition function as $(3/2)\log T$. Section \ref{sec3} is for the case where there is a gauge field present while the black hole remains static, i.e., the accelerating Reissner-Nordstr\"om black hole. In the absence of rotation, the near horizon extremal geometry is a direct product of AdS$_2$ and S$^2$ with warping factors, and the gravitational zero modes correspondingly are divided into tensor and vector modes. We introduce the finite temperature perturbation to the eigenvalue problem of the tensor and vector modes and find their contributions to the one-loop path integral of the logarithm of the partition function as $(3/2)\log T$ and $(1/2)\log T$, respectively. In addition, the gauge field fluctuation contains photon zero modes and their path integral gives another $(1/2)\log T$ term to the logarithm of the partition function. In Sec. \ref{sec4}, we consider the accelerating Kerr-Newmann black hole. The presence of rotation sets the near horizon extremal geometry as a warped and twisted product of AdS$_2$ and S$^2$. Similar to the accelerating Kerr case, the gravitational zero modes defined on the extremal background only contains tensor type modes, and their contribution to the logarithm of the partition function is $(3/2)\log T$. The gauge field fluctuation also contains photon zero modes and produce a $(1/2)\log T$ term to the logarithm of the partition function. With all the three cases analyzed, we can see a pattern that the zero modes structure is highly dependent on the near horizon extremal geometry as well as the field contents in the theory. The $\log T$ quantum correction to the logarithm of the partition function at low temperature is a universal result with coefficient $3/2$ resulting from each tensor type zero modes and $1/2$ from each vector type zero modes.

\section{Accelerating Kerr black hole}\label{sec2}
We begin with the accelerating Kerr black hole, namely, a four-dimensional rotating black hole with acceleration. This is a solution to the pure Einstein gravity, with its metric in Boyer-Lindquist coordinates taking the following form~\cite{Hong:2004dm, Griffiths:2005se, Griffiths:2005qp}
\be\label{BLg}
ds^2=-\frac{\Delta_r}{H^2\rho^2}\left(\frac{dt}{b}-a\sin^2\theta d\phi\right)^2+\frac{\rho^2}{H^2}\left(\frac{dr^2}{\Delta_r}+\frac{d\theta^2}{\Delta_{\theta}}\right)+\frac{\Delta_{\theta}a^2\sin^2\theta}{H^2\rho^2}\left(\frac{dt}{b}-\frac{r^2+a^2}{a}d\phi\right)^2\,,
\ee
where
\begin{align}
H&=1-\alpha r\cos\theta\,,~~\rho^2=r^2+a^2\cos^2\theta\,,\nn\\
\Delta_{\theta}&=1-2\alpha m\cos\theta+\alpha^2a^2\cos^2\theta\,,\\
\Delta_r&=(1-\alpha^2r^2)X\,,~~X=r^2-2mr+a^2\,.\nn
\end{align}
$m$, $a$, $\alpha$ are the black hole mass, angular momentum and acceleration parameters, respectively. In addition, $b$ is the time scaling parameter which is included for normalizing the horizon generator to obtain correct thermodynamics~\cite{Anabalon:2018qfv}. This solution of the vacuum Einstein equation characterises a pair of rotating black holes uniformly accelerating away from each other under the action of cosmic strings represented by conical singularities along the axis of symmetry. Near the north ($\theta=0$) and south ($\theta=\pi$) poles, the constant $t$, $r$, $\theta$ lines are small spatial circles with the ratio of circumference to radius given by
\be
\lim_{\theta\to0}\frac{2\pi}{|\sin\theta|}\sqrt{\frac{g_{\phi\phi}}{g_{\theta\theta}}}=2\pi\Theta_+\,,~~~~\lim_{\theta\to\pi}\frac{2\pi}{|\sin\theta|}\sqrt{\frac{g_{\phi\phi}}{g_{\theta\theta}}}=2\pi\Theta_-\,,
\ee
where $\Theta_{\pm}=1\mp2\alpha m+\alpha^2a^2$ and we set the period of the polar angle $\phi$ to be $2\pi$. Note that we can introduce a constant factor $K$ to the polar angle $\phi/K$ so that one of the conical singularities can be removed. As we will see, such a constant will be canceled in the calculation of the eigenvalues of the zero modes since the zero modes are normalized. So, here we keep the range of the polar angle as $2\pi$ for simplicity. The deficit angles $\delta_{\pm}=2\pi-2\pi\Theta_{\pm}$ at the poles are effectively induced by the cosmic strings attached on the poles with tensions $\mu_{\pm}=\delta_{\pm}/(8\pi)$. In terms of the black hole parameters, we have
\be\label{mupm}
\mu_+=\frac{1}{4}(2\alpha m-\alpha^2a^2)\,,~~~~\mu_-=\frac{1}{4}(-2\alpha m-\alpha^2a^2)\,,
\ee
and their difference is the net string tension
\be\label{munet}
\mu=\mu_+-\mu_-=m\alpha\,,
\ee
that causes the acceleration of the black hole. $\Delta_r$ is the horizon polynomial whose zeros indicate the radius of horizons. The outer and inner horizon radius $r_+$ and $r_-$ are given by
\be\label{rpm}
r_+=m+\sqrt{m^2-a^2}\,,~~~~r_-=m-\sqrt{m^2-a^2}\,,
\ee
which are zeros of $X$. The remaining positive zero point of $\Delta_r$ labels the accelerating horizon at $r_{\alpha}=1/\alpha$.

The thermodynamic quantities, for example, the mass $M$, temperature $T$, entropy $S$, angular velocity $\Omega$, angular momentum $J$ and thermodynamic lengths $\lambda_{\pm}$ conjugate to the string tensions $\mu_{\pm}$, defined with respect to the outer horizon, are given by~\cite{Anabalon:2018qfv}
\begin{align}
M&=\frac{m(1-\alpha^2a^2)}{b(1+\alpha^2a^2)}\,,~~~~T=\frac{(r_+-r_-)(1-\alpha^2r_+^2)}{4\pi b(r_+^2+a^2)}\,,~~~~S=\frac{\pi(r_+^2+a^2)}{1-\alpha^2r_+^2}\,,\nn\\
\Omega&=\Omega_H-\Omega_{\infty}=\frac{a}{b(r_+^2+a^2)}-\frac{\alpha^2a}{b(1+\alpha^2a^2)}\,,~~~~J=ma\,,\label{thermq}\\
\lambda_{\pm}&=\frac{r_+}{b(1\mp\alpha r_+)}-\frac{M}{1+\alpha^2a^2}\pm\frac{\alpha a^2}{b(1+\alpha^2a^2)}\,.\nn
\end{align}
Providing that the time scaling parameter is chosen as
\be
b=\frac{\sqrt{1-\alpha^2a^2}}{\sqrt{1+\alpha^2a^2}}\,,
\ee
the thermodynamic quantities \eqref{thermq} and the string tensions \eqref{mupm} satisfy the first law of black hole thermodynamics
\be
dM=TdS+\Omega dJ-\lambda_+\mu_+-\lambda_-\mu_-\,.
\ee

\subsection{Near extremal thermodynamics and geometry}\label{sec2.1}
We will focus on the near extremal thermodynamics where the outer and inner horizons are close to each other. In the extremal case, these two horizons are coincide, i.e., $r_+=r_-=r_0$, where $r_0$ denotes the extremal horizon radius. The accelerating horizon is still set far away from the extremal horizon. We will consider the near extremal case by introducing small deviations of $r_+$ and $r_-$ from their extremal value $r_0$.

The accelerating Kerr black hole \eqref{BLg} is parametrized by $m$, $a$, $\alpha$, so there are three independent thermodynamic quantities that can be used to describe all the other thermodynamic quantities. We take this triple as the temperature $T$, angular momentum $J$ and net string tension $\mu$. The black hole temperature $T$ in \eqref{thermq} vanishes in the extremal case. A small temperature therefore parametrizes a small deviation from extremality. The angular momentum $J_0=r_0^2$ in \eqref{thermq} and the net string tension $\mu_0=r_0\alpha$ \eqref{munet} are freely chosen as fixed at their extremal values. Thus in the near extremal case, the other thermodynamic quantities can be expanded as power series in terms of the small temperature. By using relations \eqref{rpm} and \eqref{thermq}, we find that the horizon radius $r_+$ and $r_-$ can be expanded as
\begin{align}
r_+&=r_0+\frac{4\pi r_0^2}{\sqrt{1-\mu_0^4}}T+\frac{4\pi^2r_0^3(5+3\mu_0^2)}{(1+\mu_0^2)(1-\mu_0^2)^2}T^2+\frac{32\pi^3r_0^4(4+14\mu_0^2+11\mu_0^4+3\mu_0^6)}{(1+\mu_0^2)^{\frac{5}{2}}(1-\mu_0^2)^{\frac{7}{2}}}T^3+\mathcal{O}(T^4)\,,\label{rpexpan}\\
r_-&=r_0-\frac{4\pi r_0^2}{\sqrt{1-\mu_0^4}}T-\frac{4\pi^2r_0^3(3+5\mu_0^2)}{(1+\mu_0^2)(1-\mu_0^2)^2}T^2-\frac{32\pi^3r_0^4(2+12\mu_0^2+13\mu_0^4+5\mu_0^6)}{(1+\mu_0^2)^{\frac{5}{2}}(1-\mu_0^2)^{\frac{7}{2}}}T^3+\mathcal{O}(T^4)\,,\label{rmexpan}
\end{align}
and the black hole mass $M$ and entropy $S$ take the following expansions
\begin{align}
M&=\frac{r_0\sqrt{1-\mu_0^2}}{\sqrt{1+\mu_0^2}}+\frac{4\pi^2r_0^3(1+2\mu_0^2-\mu_0^4)}{(1+\mu_0^2)^{\frac{5}{2}}(1-\mu_0^2)^{\frac{3}{2}}}T^2+\frac{32\pi^3r_0^4(1+2\mu_0^2-\mu_0^4)}{(1+\mu_0^2)^2(1-\mu_0^2)^3}T^3+\mathcal{O}(T^4)\,,\label{Mexpan}\\
S&=\frac{2\pi r_0^2}{1-\mu_0^2}+\frac{8\pi^2r_0^3\sqrt{1-\mu_0^4}}{(1-\mu_0^2)^3}T+\frac{16\pi^3r_0^4(3+9\mu_0^2+4\mu_0^4)}{(1+\mu_0^2)(1-\mu_0^2)^4}T^2+\mathcal{O}(T^3)\,.\label{Sexpan}
\end{align}
Note that the outer horizon is inside the accelerating horizon, so we have $0<\mu_0=r_0\alpha<1$. The first term on the right side of \eqref{Mexpan} is understood as the the thermal energy at extremality, while the second term is the leading thermal energy above extremality which scales quadratically with the temperature of the black hole. However, the typical energy of a quantum of Hawking radiation scales linearly with the temperature. Below the temperature where these two energy scales meet, the semiclassical analysis gives problematic low-temperature behaviors of the thermodynamic quantities since the emission of a single Hawking quantum could carry away all of the thermal energy available in the black hole system. This is known as the mass gap puzzle first discussed in~\cite{Preskill:1991tb}.

In order to resolve this puzzle, one need to go beyond the semiclassical analysis and introduce vital quantum corrections to the thermodynamics at low temperature. From the perspective of thermal partition function, quantum corrections are introduced by the Euclidean path integral over quantum fluctuations around a given semiclassical saddle-point. Since we are considering the quantum corrections for low-temperature thermodynamics, the saddle-point here should be chosen as the extremal geometry of the black hole. To obtain the extremal geometry, we perform the following scaling transformation
\be
r=r_++\frac{4\pi r_0^2}{\sqrt{1-\mu_0^4}}T(y-1)\,,~~~~t=-ib\sqrt{\frac{1+\mu_0^2}{1-\mu_0^2}}\frac{\tau}{2\pi T}\,,~~~~\phi=\frac{\varphi}{1-\mu_0^2}-i\sqrt{\frac{1+\mu_0^2}{1-\mu_0^2}}\frac{\tau}{4\pi r_0T}+i\frac{\tau}{1-\mu_0^2}\,,~~~~T\to0\,,\label{nexct}
\ee
where $\tau$ is the Euclidean time with period $2\pi$, $y\geq1$ and $\varphi$ is identified as
\be\label{pdvp}
\varphi\sim\varphi+2\pi(1-\mu_0^2)\,.
\ee
This transformation enables us to zoom into the infinite near horizon region of the near extremal accelerating Kerr black hole with a small temperature. Using the expansions \eqref{rpexpan}, \eqref{rmexpan}, and the coordinate transformation \eqref{nexct}, the black hole metric \eqref{BLg} can be written as a power series of $T$. The leading $\mathcal{O}(T^0)$ term of the metric describe the extremal geometry, which is
\be
d\bar{s}^2=\frac{r_0^2(1+\cos^2\theta)}{(1-\mu_0^2)(1-\mu_0\cos\theta)^2}\left((y^2-1)d\tau^2+\frac{dy^2}{y^2-1}\right)+\frac{r_0^2(1+\cos^2\theta)}{(1-\mu_0\cos\theta)^4}d\theta^2+\frac{4r_0^2\sin^2\theta}{(1-\mu_0^2)^2(1+\cos^2\theta)}(d\varphi-i(y-1)d\tau)^2\,,\label{g0}
\ee
and the subleading $\mathcal{O}(T)$ metric is given by
\begin{align}
\frac{\delta(ds^2)}{T}=&\frac{8\pi r_0\mu_0y\cos\theta}{\sqrt{1-\mu_0^4}(1-\mu_0\cos\theta)}d\bar{s}^2\nn\\
&+g_1(y, \theta)d\tau^2+g_2(y, \theta)dy^2+g_3(y, \theta)d\theta^2+\left(g_4(y, \theta)d\varphi+g_5(y, \theta)d\tau\right)\left(d\varphi-i(y-1)d\tau\right)\,,\label{g1}
\end{align}
where
\begin{align}
g_1(y, \theta)&=\frac{8\pi r_0^3\left[(1+\mu_0^2)(1+\cos^2\theta)-(y^2+y)(\mu_0^2+\cos^2\theta)\right](y-1)}{(1-\mu_0^2)^2\sqrt{1-\mu_0^4}(1-\mu_0\cos\theta)^2}\,,\\
g_2(y, \theta)&=\frac{8\pi r_0^3\left[-(1+\mu_0^2)(1+\cos^2\theta)+(y^2+y)(1+\mu_0^2\cos^2\theta)\right]}{(1-\mu_0^2)^2\sqrt{1-\mu_0^4}(1-\mu_0\cos\theta)^2(y^2-1)(y+1)}\,,\\
g_3(y, \theta)&=\frac{8\pi r_0^3y}{\sqrt{1-\mu_0^4}(1-\mu_0\cos\theta)^4}\,,\\
g_4(y, \theta)&=\frac{8\pi r_0^3y\sin^2(2\theta)}{(1-\mu_0^2)^2\sqrt{1-\mu_0^4}(1+\cos^2\theta)^2}\,,\\
g_5(y, \theta)&=\frac{8\pi ir_0^3\sin^2\theta}{(1-\mu_0^2)^3\sqrt{1-\mu_0^4}(1-\mu_0\cos\theta)^2(1+\cos^2\theta)^2}\Big[-(1-\mu_0^2)^2(y^2-1)(1+\cos^2\theta)^2\nn\\
&+(1-\mu_0\cos\theta)^2\left(2y^2(1-\mu_0^2)-5-3\mu_0^2-\left((2y^2-4y)(1-\mu_0^2)+5+3\mu_0^2\right)\cos^2\theta\right)\Big]\,.
\end{align}
The metric \eqref{g0} describes the near horizon extremal geometry of the accelerating Kerr black hole, which is a warped and twisted product of AdS$_2$ and S$^2$. The near horizon near extremal geometry $d\bar{s}^2+\delta(ds^2)$ is obtained by adding the first-order low-temperature perturbation \eqref{g1}. In the following discussion, the near extremal configuration will be used as the perturbed background in the Euclidean path integral and determines the leading quantum corrections to the thermodynamic quantities of the accelerating Kerr black hole.

\subsection{Euclidean path integral near extremality and zero modes}\label{sec2.2}
The Euclidean path integral formally determines the partition function of the theory by integrating over all possible Euclidean configurations. For the Einstein gravity in four dimensions, the partition function is given by
\be
Z=\int[Dg]\mathrm{exp}(-I_\mathrm{E}[g])\,,~~~~I_\mathrm{E}[g]=-\frac{1}{16\pi}\int_\mathcal{M}d^4x\sqrt{g}R+I_{\mathrm{B.}}+I_{\mathrm{G. F.}}\,,
\ee
where $R$ is the Ricci scalar. The boundary term $I_{\mathrm{B.}}$ is included for the consistency of the variational principle under certain boundary conditions and the gauge fixing term $I_{\mathrm{G. F.}}$ is introduced for removing part of the gauge redundancy of the theory.

To evaluate the partition function, one can expand the metric around a saddle-point $\bar{g}$
\be
g=\bar{g}+h\,,
\ee
where $h$ is the quantum fluctuation of the metric. Then the one-loop approximation of the partition function can be written as
\be\label{Zexpan}
Z\approx\mathrm{exp}(-I_E[\bar{g}])\int[Dh]\mathrm{exp}\left(-\int d^4x\sqrt{\bar{g}}h^*\bar{\Delta}[\bar{g}]h\right)\,.
\ee
where $*$ denotes complex conjugation. The first-order term vanishes as the classical equations of motion are satisfied. The linearized kinetic operator $\bar{\Delta}[\bar{g}]$ in \eqref{Zexpan} depends on the saddle-point as well as the gauge fixing condition. We will use the harmonic gauge condition in the following discussion with the Lagrange density of the gauge fixing term given by
\be\label{hg}
\mathcal{L}_{\mathrm{G. F.}}=\frac{1}{32\pi}\bar{g}_{\mu\nu}\left(\bar{\nabla}_{\alpha}h^{\alpha\mu}-\frac{1}{2}\bar{\nabla}^{\mu}h^{\alpha}_{~\alpha}\right)\left(\bar{\nabla}_{\beta}h^{\beta\nu}-\frac{1}{2}\bar{\nabla}^{\nu}h^{\beta}_{~\beta}\right)\,.
\ee
For the harmonic gauge, the linearized kinetic operator acting on $h_{\mu\nu}$ takes the form~\cite{Christensen:1979iy}
\begin{align}
h^*_{\alpha\beta}\bar{\Delta}^{\alpha\beta,\mu\nu}[\bar{g}]h_{\mu\nu}=&-\frac{1}{16\pi}h^*_{\alpha\beta}\Bigg(\frac{1}{4}\bar{g}^{\alpha\mu}\bar{g}^{\beta\nu}\bar{\Box}-\frac{1}{8}\bar{g}^{\alpha\beta}\bar{g}^{\mu\nu}\bar{\Box}+\frac{1}{2}\bar{R}^{\alpha\mu\beta\nu}+\frac{1}{2}\bar{R}^{\alpha\mu}\bar{g}^{\beta\nu}-\frac{1}{2}\bar{R}^{\alpha\beta}\bar{g}^{\mu\nu}\nn\\
&-\frac{1}{4}\bar{R}\bar{g}^{\alpha\mu}\bar{g}^{\beta\nu}+\frac{1}{8}\bar{R}\bar{g}^{\alpha\beta}\bar{g}^{\mu\nu}\Bigg)h_{\mu\nu}\,.\label{ko}
\end{align}
Once the saddle-point $\bar{g}$ is given, the one-loop partition function \eqref{Zexpan} is determined by the eigenvalues of the linearized kinetic operator $\bar{\Delta}[\bar{g}]$. We will use the saddle-point \eqref{g0}, which is the near horizon extremal geometry and satisfies the vacuum Einstein equation, to evaluate the one-loop partition function of the extremal accelerating Kerr black hole. The resultant quantum corrections to the black hole partition function help to resolve the mass gap puzzle discussed in Sec. \ref{sec2.1}.

The modes $h_{\mu\nu}$ with vanishing eigenvalues of the operator $\bar{\Delta}[\bar{g}]$, which are called zero modes, play crucial roles in the one-loop approximation of the path integral. As we will see in the next subsection, zero modes contribute leading quantum corrections to the partition function in the low-temperature limit. Zero modes are usually associated to the background symmetry that are not removed by the gauge fixing. In the harmonic gauge \eqref{hg}, zero modes are necessary to satisfy the transverse and traceless conditions and thus are generated by large diffeomorphisms. In other words, a zero mode can be written as a Lie derivative of the background geometry along a certain vector field $\xi$ which has nontrivial behavior at the boundary. For the saddle-point \eqref{g0}, we take the following ansatz for the vector field
\be
\xi=e^{in\tau}\left(f_{\tau}(y)\frac{\p}{\p\tau}+f_y(y)\frac{\p}{\p y}+f_{\varphi}(y)\frac{\p}{\p\varphi}\right)\,,~~~~n\in\mathbb{Z}\,.\label{xiansatz}
\ee
The Euclidean time $\tau$ has period $2\pi$ due to the horizon smoothness condition, so here $n$ takes integer values. The possible zero modes thus can be written as $h_{\mu\nu}=\mathcal{L}_{\xi}\bar{g}_{\mu\nu}$ up to normalization. To find their specific forms, we need to solve the functions $f_{\tau}(y), f_y(y), f_{\varphi}(y)$ by using the transverse and traceless conditions or the zero eigenvalue condition of the operator defined in \eqref{ko}. We start from the traceless condition $h^{\alpha}_{~\alpha}=0$, from which it is easy to find
\be\label{ftau}
f_{\tau}(y)=\frac{if_y'(y)}{n}\,,~~~~n\neq0\,.
\ee
Next, recalling \eqref{ftau} and using the operator \eqref{ko}, we find that
\be
\frac{1-\mu_0}{1+\mu_0}\p_{\theta}\left(\bar{\Delta}[\bar{g}]h\right)_{y\theta}|_{\theta=0}+\frac{1}{2(1+\mu_0^2)}\p_{\theta}\left(\bar{\Delta}[\bar{g}]h\right)_{y\theta}|_{\theta=\frac{\pi}{2}}\propto f_y(y)-(y-1)f_y'(y)-nf_{\varphi}(y)\,.
\ee
The zero eigenvalue condition is satisfied given that
\be\label{fvarphi}
f_{\varphi}(y)=\frac{f_y(y)-(y-1)f_y'(y)}{n}\,,~~~~n\neq0\,.
\ee
Finally, recalling \eqref{ftau}, \eqref{fvarphi} and using the transverse condition $\bar{\nabla}_{\alpha}h^{\alpha}_{~\mu}=0$, we have
\be
\bar{\nabla}_{\alpha}h^{\alpha}_{~y}\propto(y^2-1)f_y''(y)+2yf_y'(y)-\left(2+\frac{n^2}{y^2-1}\right)f_y(y)=0\,,
\ee
which is a second-order differential equation of $f_y(y)$ and is solved by the following general solution
\be
f_y(y)=C_1\left(\frac{y-1}{y+1}\right)^{\frac{n}{2}}(n+y)+C_2\left(\frac{y-1}{y+1}\right)^{-\frac{n}{2}}(-n+y)\,,~~~~n\neq0\,,
\ee
where $C_1$ and $C_2$ are arbitrary constants. The above solution is valid for any integer $n\neq0$. However, one should set $C_2=0$ for $n>0$ and $C_1=0$ for $n<0$ for the sake of the regularity of the vector field $\xi$ on the horizon $y=1$. So equivalently, one can rewrite the regular solution as
\be\label{fy}
f_y(y)=Cn\left(\frac{y-1}{y+1}\right)^{\frac{|n|}{2}}(|n|+y)\,,~~~~n\in\mathbb{Z}\,,
\ee
with $C$ bing an arbitrary constant. Note that an additional $n$ factor is incorporated in \eqref{fy} for recovering the $n=0$ case. Thus, the vector field $\xi$ can be written as
\be
\xi=Ce^{in\tau}\left(\frac{y-1}{y+1}\right)^{\frac{|n|}{2}}\left(\frac{i\left(|n|(|n|+y)+y^2-1\right)}{y^2-1}\frac{\p}{\p\tau}+n(|n|+y)\frac{\p}{\p y}+\frac{|n|-n^2+y+1}{y+1}\frac{\p}{\p\varphi}\right)\,,\label{xi}
\ee
with integer valued $n$. These vector fields are large diffeomorphisms since they have nontrivial behaviors at the asymptotic boundary $y\to+\infty$,
\be
\xi\to Ce^{in\tau}\left(i\frac{\p}{\p\tau}+ny\frac{\p}{\p y}+\frac{\p}{\p\varphi}\right)\,,~~~~n\in\mathbb{Z}\,,
\ee
which correspond to the boundary time reparametrizations with fixed boundary curve length~\cite{Maldacena:2016upp}.

The zero modes generated by \eqref{xi} then take the following forms
\begin{align}
h_{\mu\nu}^{(n)}dx^{\mu}dx^{\nu}=&e^{in\tau}\frac{\sqrt{3|n|(n^2-1)}(1-\mu_0^2)(1+\cos^2\theta)}{8\pi\sqrt{1+\mu_0^2}(1-\mu_0\cos\theta)^2(y^2-1)}\left(\frac{y-1}{y+1}\right)^{\frac{|n|}{2}}\nn\\
&\times\left(-(y^2-1)d\tau^2+2i\frac{|n|}{n}d\tau dy+\frac{1}{y^2-1}dy^2\right)\,,~~~~|n|\geq2\,,\label{zeromodes}
\end{align}
where the arbitrary constant $C$ in \eqref{xi} is chosen so that $h_{\mu\nu}^{(n)}$ are orthonormalized
\be\label{onc}
\int d^4x\sqrt{\bar{g}}h_{\alpha\beta}^{(m)*}\bar{g}^{\alpha\mu}\bar{g}^{\beta\nu}h_{\mu\nu}^{(n)}=\delta^{mn}\,.
\ee
Note that the $n=0, \pm1$ zero modes vanish due to the fact that their corresponding vector fields $\xi$ are Killing vectors of $\bar{g}$ thus do not contribute. One can directly check that the fluctuations \eqref{zeromodes} satisfy $\left(\bar{\Delta}[\bar{g}]h^{(n)}\right)_{\mu\nu}=0$.

However, due to the vanishing of the eigenvalues and the infinite volumes of the zero modes, a direct calculation of the path integral over the zero modes leads to an infrared divergence of the one-loop partition function
\be
Z\propto\int_{\mathrm{zero~modes}}[Dh]\to\infty\,.
\ee
One way to regularize the one-loop partition function is to deform the extremal background by turning on slightly the temperature of the black hole, which in turn will corrects slightly the eigenvalues of the zero modes. In the next subsection, we will use the near extremal configuration given by \eqref{g0} and \eqref{g1} to calculate the leading corrections to the eigenvalues of the zero modes. We will see that these slightly corrected nonvanishing eigenvalues alter the partition function dramatically with a leading low-temperature effect.

\subsection{Quantum corrections to the entropy}\label{sec2.3}
Since the one-loop partition \eqref{Zexpan} is a standard Gaussian integral, the result is controlled by the functional determinant and thus the eigenvalues of the linearized kinetic operator \eqref{ko}. Denoting $h^{(n)}$ and $\bar{\Lambda}_n$ as the eigenstates and eigenvalues of $\bar{\Delta}[\bar{g}]$, respectively, i.e., $\bar{\Delta}[\bar{g}]h^{(n)}=\bar{\Lambda}_nh^{(n)}$, then the one-loop partition function scales as $Z\sim\prod_n\bar{\Lambda}_n^{-1/2}$ up to an overall factor. It is clear that for the zero modes with $\bar{\Lambda}_n=0$, we have $Z\to\infty$.

Now let us regularize the divergence in the one-loop partition function by departing slightly the eigenvalues of the zero modes from zeros. The way is to heat up the geometry by adding the $\mathcal{O}(T)$ perturbation \eqref{g1} to the extremal geometry \eqref{g0}. Such a deformation of the background geometry
\be
\bar{g}\to\bar{g}+\delta g
\ee
will change the eigenspectrum problem to
\be\label{Dperturb}
\left(\bar{\Delta}[\bar{g}]+\delta\Delta[\bar{g}, \delta g]\right)\left(h^{(n)}+\delta h^{(n)}\right)=\left(\bar{\Lambda}_n+\delta\Lambda_n\right)\left(h^{(n)}+\delta h^{(n)}\right)\,,
\ee
where $\delta\Delta[\bar{g}, \delta g]$ is the perturbed linearized kinetic operator and $(\delta h^{(n)}, \delta\Lambda_n)$ is the perturbed eigenspectrum. And it will also change the scaling of the one-loop partition function to
\be
Z\sim\prod_n\left(\bar{\Lambda}_n+\delta\Lambda_n\right)^{-\frac{1}{2}}\,,~~~~\log Z=-\frac{1}{2}\sum_n\log\left(\bar{\Lambda}_n+\delta\Lambda_n\right)+\cdots\,.
\ee
Note that $\delta\Lambda_n$ are small corrections to the eigenvalues which depend linearly on $T$, for nonzero modes with $\bar{\Lambda}_n\neq0$, the corrections are purely polynomial in $T$,
\be
\log Z=-\frac{1}{2}\sum_n\log\bar{\Lambda}_n-\frac{1}{2}\sum_n\frac{\delta\Lambda_n(T)}{\bar{\Lambda}_n}+\frac{1}{4}\sum_n\frac{\delta\Lambda_n(T)^2}{\bar{\Lambda}_n^2}+\cdots\,.
\ee
While for zero modes with $\bar{\Lambda}_n=0$, small $\delta\Lambda_n$ will produce leading $\log T$ correction to the entropy
\be\label{dlogZ}
\delta\log Z=-\frac{1}{2}\sum_n\log\left(\delta\Lambda_n(T)\right)\,,
\ee
and correspondingly produce $\mathcal{O}(T)$ correction to the black hole mass at low temperature. So zero modes are the essential objects to resolve the mass gap puzzle addressed in Sec. \ref{sec2.1}.

The eigenvalue corrections $\delta\Lambda_n$ for the zero modes can be figured out by using perturbation theory. Taking the inner product on both sides of \eqref{Dperturb} with $h^{(m)}$, we have
\be\label{dLambda}
\delta\Lambda_n=\langle h^{(n)}|\delta\Delta[\bar{g}, \delta g]h^{(n)}\rangle=\int d^4x\sqrt{\bar{g}}h_{\alpha\beta}^{(n)*}\delta\Delta[\bar{g}, \delta g]^{\alpha\beta,\mu\nu}h_{\mu\nu}^{(n)}\,,
\ee
where we used the fact that $\bar{\Delta}[\bar{g}]$ is self-adjoint, i.e., $\langle h^{(m)}|\bar{\Delta}[\bar{g}]\delta h^{(n)}\rangle=\langle \bar{\Delta}[\bar{g}]h^{(m)}|\delta h^{(n)}\rangle$ and $h^{(n)}$ are orthonormalized. It is worthwhile to mention that the conical singularities induced by the improper range of the polar angle $\phi$ are inherited to the near horizon extremal geometry \eqref{g0} through \eqref{pdvp}. However this effect will not affect the calculation of the eigenvalue corrections \eqref{dLambda}. The range of the polar angle $\varphi$ is canceled given that the zero modes satisfying the orthonormal condition \eqref{onc}. The integration in \eqref{dLambda} is highly nontrivial. One way to carry out the integrand is to substitute the full perturbed metric $\bar{g}+\delta g$ into the linearized kinetic operator \eqref{ko} and keep the leading $\mathcal{O}(T)$ term. The variation of the operator itself is intractable but it simplifies dramatically when contracting with the zero modes. Here we print the result we find
\be\label{dLambdares}
\delta\Lambda_n=\frac{3|n|(1-\mu_0^2)^{\frac{3}{2}}}{64r_0\sqrt{1+\mu_0^2}}T+\mathcal{O}(T^2)\,,~~~~|n|\geq2\,.
\ee
Thus, from \eqref{dlogZ}, \eqref{dLambdares} and using the zeta function regularization
\be
\prod_{n\geq2}\frac{\rho}{n}=\frac{1}{\rho^{\frac{3}{2}}\sqrt{2\pi}}\,,
\ee
for real $\rho$, we find the contribution of the zero modes to the near extremal entropy
\begin{align}
\delta\log Z&=-\frac{1}{2}\sum_{|n|\geq2}\log\left(\delta\Lambda_n\right)=-\log\prod_{n\geq2}\left(\delta\Lambda_n\right)\nn\\
&=\frac{3}{2}\log\left(\frac{T}{T_q}\right)+\log\left(\frac{3}{512}\sqrt{\frac{3}{2\pi}}\right)+\cdots\,,
\end{align}
where
\be\label{Tq}
T_q=\frac{r_0\sqrt{1+\mu_0^2}}{(1-\mu_0^2)^{\frac{3}{2}}}\,.
\ee
$T_q$ can be viewed as the critical temperature below which the low-temperature quantum corrections become important. From \eqref{Tq}, it is clear that the critical temperature rises as the acceleration of the black hole increases. However, this is not necessarily indicating that the high temperature quantum corrections are significant for a black hole with large acceleration. The mechanism presented here is working for the first-order low-temperature perturbation theory. The corrections from the levels that beyond the first order are still unknown at present stage.

Taking into account the contribution from the zero modes, the near extremal partition function can be approximately written as
\be
Z(T)\approx T^{\frac{3}{2}}e^{S_0}+\mathrm{higher~order~terms}\,.\label{ZT}
\ee
Therefore, the entropy and mass of the near extremal accelerating Kerr black hole are shifted according to
\begin{align}
&S=S_0+S'+\mathcal{O}(T)\,,\mkern-120mu &&S'=\frac{3}{2}\log T\,,\\
&M=M_0+M'+\mathcal{O}(T^2)\,,\mkern-120mu &&M'=\int TdS'=\frac{3}{2}T\,,\label{MlinearT}
\end{align}
where $S_0$ and $M_0$ are the extremal entropy and mass listed as the first terms on the right sides of \eqref{Sexpan} and \eqref{Mexpan}, respectively. So taking into account the quantum corrections at low temperature, the black hole thermal energy above extremality \eqref{MlinearT} scales linearly with the temperature. This in turn relieve the tension between the thermal energy available for the black hole and the energy carried away by the Hawking radiation at sufficiently low temperature.

\section{Accelerating Reissner-Nordstr\"om black hole}\label{sec3}
We next turn to the accelerating Reissner-Nordstr\"om black hole. Compered to the previous case, the absence of rotation and the presence of the gauge field both give rise to a richer zero modes structure. The charged C-metric and the electric gauge filed of the accelerating Reissner-Nordstr\"om black hole in the Boyer-Lindquist coordinate system take the following forms~\cite{Hong:2003gx, Griffiths:2005se, Griffiths:2005qp}
\begin{align}
ds^2&=-\frac{\Delta_r}{H^2r^2b^2}dt^2+\frac{r^2}{H^2}\left(\frac{dr^2}{\Delta_r}+\frac{d\theta^2}{\Delta_{\theta}}\right)+\frac{\Delta_\theta r^2\sin^2\theta}{H^2}d\phi^2\,,\label{BLgARN}\\
A&=-\frac{q}{rb}dt+\Phi dt\,,\label{AARN}
\end{align}
where
\begin{align}
H&=1-\alpha r\cos\theta\,,\nn\\
\Delta_{\theta}&=1-2\alpha m\cos\theta+\alpha^2q^2\cos^2\theta\,,\\
\Delta_r&=(1-\alpha^2r^2)X\,,~~X=r^2-2mr+q^2\,.\nn
\end{align}
$m$, $q$, $\alpha$ are the black hole mass, electric charge and acceleration parameters, respectively. $b$ is the time scaling parameter introduced for achieving consistent black hole thermodynamics~\cite{Anabalon:2018qfv}. $\Phi$ is the electric potential at the horizon, which ensures that the electric gauge field $A$ vanishes on the horizon. The accelerating of the charged black hole is induced by the cosmic string attached on the north ($\theta=0$) and south ($\theta=\pi$) poles. The string tensions are represented by the conical defects at each poles. Following the same argument in Sec. \ref{sec2}, the string tensions at the north $\mu_+$ and south $\mu_-$ poles can be expressed as
\be\label{mupmARN}
\mu_+=\frac{1}{4}(2\alpha m-\alpha^2q^2)\,,~~~~\mu_-=\frac{1}{4}(-2\alpha m-\alpha^2q^2)\,,
\ee
where the period of the polar angle $\phi$ is set to be $2\pi$. Their difference gives the net string tension
\be\label{munetARN}
\mu=\mu_+-\mu_-=m\alpha\,,
\ee
that cause the acceleration of the black hole. The outer and inner horizon radius $r_+$ and $r_-$ are given by
\be\label{rpmARN}
r_+=m+\sqrt{m^2-q^2}\,,~~~~r_-=m-\sqrt{m^2-q^2}\,,
\ee
and the accelerating horizon is still fixed at $r_{\alpha}=1/\alpha$. The thermodynamic quantities: mass $M$, temperature $T$, entropy $S$, electric potential $\Phi$, electric charge $Q$ and thermodynamic lengths $\lambda_{\pm}$ conjugate to the string tensions $\mu_{\pm}$, formulated with respect to the outer horizon, are given by~\cite{Anabalon:2018qfv}
\begin{align}
M&=\frac{m}{b}\,,~~~~T=\frac{(r_+-r_-)(1-\alpha^2r_+^2)}{4\pi br_+^2}\,,~~~~S=\frac{\pi r_+^2}{1-\alpha^2r_+^2}\,,\nn\\
\Phi&=\frac{q}{br_+}\,,~~~~Q=q\,,~~~~\lambda_{\pm}=\frac{r_+}{b(1\mp\alpha r_+)}-\frac{M}{1+\alpha^2q^2}\,.\label{thermqARN}
\end{align}
The above thermodynamic quantities are well defined and the thermodynamic first law holds, i.e.,
\be
dM=TdS+\Phi dQ-\lambda_+\mu_+-\lambda_-\mu_-\,,
\ee
given that
\be
b=\sqrt{1+\alpha^2q^2}\,.
\ee

\subsection{Near extremal thermodynamics and geometry}\label{sec3.1}
The near extremal thermodynamics fucus on the situation that the two horizon radius $r_+$ and $r_-$ depart a little bit from their extremal value $r_0$. Similar to the rotating case, there are three independent thermodynamic quantities which can be chosen as the temperature $T$, electric charge $Q$ and net string tension $\mu$. In the near extremal expansions, we take the temperature as a small parameter and keep the electric charge $Q_0=r_0$ and the net string tension $0<\mu_0=r_0\alpha<1$ fixed at their extremal values.

By using relations \eqref{rpmARN} and \eqref{thermqARN}, we find the following near extremal expansions for the horizon radius $r_+$ and $r_-$
\begin{align}
r_+&=r_0+\frac{2\pi r_0^2\sqrt{1+\mu_0^2}}{1-\mu_0^2}T+\frac{2\pi^2r_0^3(5+4\mu_0^2-\mu_0^4)}{(1-\mu_0^2)^3}T^2+\frac{64\pi^3r_0^4(1+\mu_0^2)^{\frac{3}{2}}}{(1-\mu_0^2)^5}T^3+\mathcal{O}(T^4)\,,\label{rpexpanARN}\\
r_-&=r_0-\frac{2\pi r_0^2\sqrt{1+\mu_0^2}}{1-\mu_0^2}T-\frac{2\pi^2r_0^3(3+4\mu_0^2+\mu_0^4)}{(1-\mu_0^2)^3}T^2-\frac{32\pi^3r_0^4(1+\mu_0^2)^{\frac{5}{2}}}{(1-\mu_0^2)^5}T^3+\mathcal{O}(T^4)\,,\label{rmexpanARN}
\end{align}
and the black hole mass $M$ and entropy $S$ can be expanded as
\begin{align}
M&=\frac{r_0}{\sqrt{1+\mu_0^2}}+\frac{2\pi^2r_0^3\sqrt{1+\mu_0^2}}{(1-\mu_0^2)^2}T^2+\frac{16\pi^3r_0^4(1+\mu_0^2)}{(1-\mu_0^2)^4}T^3+\mathcal{O}(T^4)\,,\label{MexpanARN}\\
S&=\frac{\pi r_0^2}{1-\mu_0^2}+\frac{4\pi^2r_0^3\sqrt{1+\mu_0^2}}{(1-\mu_0^2)^3}T+\frac{8\pi^3r_0^4(3+4\mu_0^2+\mu_0^4)}{(1-\mu_0^2)^5}T^2+\mathcal{O}(T^3)\,.\label{SexpanARN}
\end{align}
From \eqref{MexpanARN}, it is clear that the thermal energy above extremality scales quadratically with the temperature, so there is still a mass gap between the thermal energy and the energy carried away by the Hawking radiation which scales linearly with the temperature. Quantum corrections are thus necessary to be included to fill the gap.

The quantum corrections near extremality are introduced by using the Euclidean path integral with the saddle-point chosen as the extremal configuration. To find the near horizon near extremal configuration, we therefore perform the following coordinate transformation
\be
r=r_++\frac{2\pi r_0^2\sqrt{1+\mu_0^2}}{1-\mu_0^2}T(y-1)\,,~~~~t=-i\frac{b}{\sqrt{1+\mu_0^2}}\frac{\tau}{2\pi T}\,,~~~~T\to0\,,\label{nexctARN}
\ee
where $\tau$ is the Euclidean time with period $2\pi$ and $y\geq1$. This infinite scaling transformation enables us to zoom into the near horizon region of the near extremal accelerating Reissner-Nordstr\"om black hole with a small temperature. Using the expansions \eqref{rpexpanARN}, \eqref{rmexpanARN} and the transformation \eqref{nexctARN}, we find the leading $\mathcal{O}(T^0)$ terms of the metric \eqref{BLgARN} and the gauge field \eqref{AARN}
\begin{align}
d\bar{s}^2&=\frac{r_0^2}{(1-\mu_0^2)(1-\mu_0\cos\theta)^2}\left((y^2-1)d\tau^2+\frac{dy^2}{y^2-1}\right)+\frac{r_0^2}{(1-\mu_0\cos\theta)^4}d\theta^2+r_0^2\sin^2\theta d\phi^2\,,\label{g0ARN}\\
\bar{A}&=-\frac{ir_0(y-1)}{1-\mu_0^2}d\tau\,,\label{A0ARN}
\end{align}
and their subleading $\mathcal{O}(T)$ terms
\begin{align}
\frac{\delta(ds^2)}{T}&=\frac{4\pi r_0\mu_0\sqrt{1+\mu_0^2}y\cos\theta}{(1-\mu_0^2)(1-\mu_0\cos\theta)}d\bar{s}^2+g_1(y, \theta)d\tau^2+g_2(y, \theta)dy^2+g_3(y, \theta)d\theta^2+g_4(y, \theta)d\phi^2\,,\label{g1ARN}\\
\frac{\delta A}{T}&=\frac{2ir_0^2\sqrt{1+\mu_0^2}(y^2-1)}{(1-\mu_0^2)^2}d\tau\label{A1ARN}
\end{align}
where
\begin{align}
g_1(y, \theta)&=-\frac{4\pi r_0^3\sqrt{1+\mu_0^2}(y^3-3y+2)}{(1-\mu_0^2)^3(1-\mu_0\cos\theta)^2}\,,\\
g_2(y, \theta)&=\frac{4\pi r_0^3\sqrt{1+\mu_0^2}(y+2)}{(1-\mu_0^2)^3(1-\mu_0\cos\theta)^2(y+1)^2}\,,\\
g_3(y, \theta)&=\frac{4\pi r_0^3\sqrt{1+\mu_0^2}y}{(1-\mu_0^2)(1-\mu_0\cos\theta)^4}\,,\\
g_4(y, \theta)&=\frac{4\pi r_0^3\sqrt{1+\mu_0^2}y\sin^2\theta}{1-\mu_0^2}\,.
\end{align}
The metric \eqref{g0ARN} characterize the near horizon extremal geometry of the accelerating Reissner-Nordstr\"om black hole, which is a direct product of AdS$_2$ and S$^2$ with warping factors. Equation \eqref{A0ARN} is the extremal gauge field, which vanishes explicitly on the horizon $y=1$. When combined with the subleading terms \eqref{g1ARN} and \eqref{A1ARN}, we get the near horizon near extremal configurations $d\bar{s}^2+\delta(ds^2)$ and $\bar{A}+\delta(A)$ up to $\mathcal{O}(T)$.

\subsection{Euclidean path integral near extremality and zero modes}\label{sec3.2}
For the Einsten-Maxwell theory in four dimensions, the partition function is given by the following Euclidean path integral
\be
Z=\int[Dg]\mathrm{exp}(-I_\mathrm{E}[g, A])\,,~~~~I_\mathrm{E}[g, A]=-\frac{1}{16\pi}\int_\mathcal{M}d^4x\sqrt{g}(R-F^2)+I_{\mathrm{B.}}+I_{\mathrm{G. F.}}\,,
\ee
where $R$ is the Ricci scalar, $F^2=F_{\mu\nu}F^{\mu\nu}$ and $F_{\mu\nu}$ is the field strength of the gauge field. $I_{\mathrm{B}.}$ is the boundary term and $I_{\mathrm{G. F.}}$ is the gauge fixing term.

The one-loop approximation of the partition function includes contributions from both the metric and the gauge field quantum fluctuations. Expand the metric and the gauge field as
\be
g=\bar{g}+h\,,~~~~A=\bar{A}+\frac{1}{2}a\,,
\ee
where $\bar{g}$ and $\bar{A}$ are saddle-points of the metric and the gauge field, respectively, $h$ and $a$ are their quantum fluctuations, then the one-loop partition function can be written as
\be
Z\approx\mathrm{exp}(-I_E[\bar{g}, \bar{A}])\int[Dh][Da]\mathrm{exp}\left[-\int d^4x\sqrt{\bar{g}}\left(h^*\bar{\Delta}[\bar{g}, \bar{A}]h+a^*\bar{P}[\bar{g}, \bar{A}]a+\left(h^*\bar{O}_{\mathrm{int}}[\bar{g}, \bar{A}]a+\mathrm{h. c.}\right)\right)\right]\,,\label{ZexpanARN}
\ee
where $*$ denotes complex conjugation. The explicit forms of the one-loop kinetic operators $\bar{\Delta}[\bar{g}, \bar{A}]$, $\bar{P}[\bar{g}, \bar{A}]$ and $\bar{O}_{\mathrm{int}}[\bar{g}, \bar{A}]$ in \eqref{ZexpanARN} are sensitive to the gauge fixing condition. In the following discussion, we will use the harmonic gauge for the metric fluctuation and the Lorentz gauge for the gauge field one, i.e., the Lagrange density of the gauge fixing term is given by
\be
\mathcal{L}_{\mathrm{G. F.}}=\mathcal{L}_{\mathrm{metric}}+\mathcal{L}_{\mathrm{gauge}}\,,
\ee
where
\begin{align}
\mathcal{L}_{\mathrm{metric}}&=\frac{1}{32\pi}\bar{g}_{\mu\nu}\left(\bar{\nabla}_{\alpha}h^{\alpha\mu}-\frac{1}{2}\bar{\nabla}^{\mu}h^{\alpha}_{~\alpha}\right)\left(\bar{\nabla}_{\alpha}h^{\alpha\nu}-\frac{1}{2}\bar{\nabla}^{\nu}h^{\alpha}_{~\alpha}\right)\,,\\
\mathcal{L}_{\mathrm{gauge}}&=\frac{1}{32\pi}\left(\bar{\nabla}^{\mu}a_{\mu}\right)^2\,.
\end{align}
With these gauge fixings, the one-loop kinetic operators for the Einstein-Maxwell theory take the following forms~\cite{Blacker:2025zca}
\begin{align}
h^*_{\alpha\beta}\bar{\Delta}^{\alpha\beta,\mu\nu}[\bar{g}, \bar{A}]h_{\mu\nu}=&-\frac{1}{16\pi}h^*_{\alpha\beta}\Bigg(\frac{1}{4}\bar{g}^{\alpha\mu}\bar{g}^{\beta\nu}\bar{\Box}-\frac{1}{8}\bar{g}^{\alpha\beta}\bar{g}^{\mu\nu}\bar{\Box}+\frac{1}{2}\bar{R}^{\alpha\mu\beta\nu}+\frac{1}{2}\bar{R}^{\alpha\mu}\bar{g}^{\beta\nu}-\frac{1}{2}\bar{R}^{\alpha\beta}\bar{g}^{\mu\nu}\nn\\
&-\frac{1}{4}\bar{R}\bar{g}^{\alpha\mu}\bar{g}^{\beta\nu}+\frac{1}{8}\bar{R}\bar{g}^{\alpha\beta}\bar{g}^{\mu\nu}+\frac{1}{4}\bar{F}^2\bar{g}^{\alpha\mu}\bar{g}^{\beta\nu}-\frac{1}{8}\bar{F}^2\bar{g}^{\alpha\beta}\bar{g}^{\mu\nu}-\bar{F}^{\alpha\mu}\bar{F}^{\beta\nu}\nn\\
&-2\bar{F}^{\alpha\gamma}\bar{F}^{\mu}_{~\gamma}\bar{g}^{\beta\nu}+\bar{F}^{\alpha\gamma}\bar{F}^{\beta}_{~\gamma}\bar{g}^{\mu\nu}\Bigg)h_{\mu\nu}\,,\label{koDARN}\\
a^*_{\mu}\bar{P}^{\alpha\beta}[\bar{g}, \bar{A}]a_{\mu}=&-\frac{1}{32\pi}a^*_{\mu}\left(\bar{g}^{\mu\nu}\bar{\Box}-\bar{R}^{\mu\nu}\right)a_{\nu}\,,\label{koPARN}\\
h^*_{\alpha\beta}\bar{O}_{\mathrm{int}}^{\alpha\beta\mu}[\bar{g}, \bar{A}]a_{\mu}=&\frac{1}{16\pi}h^*_{\alpha\beta}\left(4g^{\alpha[\mu}\bar{F}^{\nu]\beta}+\bar{F}^{\mu\nu}\bar{g}^{\alpha\beta}\right)\bar{\nabla}_{\mu}a_{\nu}\,.\label{koMARN}
\end{align}
The one-loop partition function \eqref{ZexpanARN} therefore scales as the functional determinant of the matrix operator
\be\label{boARN}
Z\sim\mathrm{det}\left(
                   \begin{array}{cc}
                     \bar{\Delta}[\bar{g}, \bar{A}] & \bar{O}_{\mathrm{int}}[\bar{g}, \bar{A}] \\
                     \bar{O}^{\dagger}_{\mathrm{int}}[\bar{g}, \bar{A}] & \bar{P}[\bar{g}, \bar{A}] \\
                   \end{array}
                 \right)^{-\frac{1}{2}}\,.
\ee
which necessitates the joint diagonalization of the operators $\bar{\Delta}[\bar{g}, \bar{A}]$, $\bar{P}[\bar{g}, \bar{A}]$, and $\bar{O}_{\mathrm{int}}[\bar{g}, \bar{A}]$. Inspired by the previous accelerating Kerr black hole case, we will focus on the path integral over zero modes of the saddle-points \eqref{g0ARN} and \eqref{A0ARN}, which hopefully produces the leading quantum correction to the partition function at low temperature. Terminologically, the zero modes of the matrix operator in \eqref{boARN} can be divided into graviton zero modes and photon zero modes.

The graviton zero modes, which satisfy $\left(\bar{\Delta}[\bar{g}, \bar{A}]h\right)_{\mu\nu}=0$, are generated by the large diffeomorphisms that are not removed by the harmonic gauge. We take the following ansatz for the vector fields associated to the large diffeomorphisms
\be
\xi=e^{in\tau}\left(f_{\tau}(y)\frac{\p}{\p\tau}+f_y(y)\frac{\p}{\p y}+f_{\phi}(y)\frac{\p}{\p\phi}\right)\,,~~~~n\in\mathbb{Z}\,.
\ee
Given the extremal geometry \eqref{g0ARN} and the functions $f_{\tau}(y)$, $f_y(y)$, $f_{\phi}(y)$, the graviton zero modes are then explicitly determined by $h_{\mu\nu}=\mathcal{L}_{\xi}\bar{g}_{\mu\nu}$ up to normalization. In the harmonic gauge, the traceless condition $h^{\alpha}_{~\alpha}=0$ of the zero modes gives
\be\label{ftauARN}
f_{\tau}(y)=\frac{if_y'(y)}{n}\,,~~~~n\neq0\,.
\ee
Combined with the above relation, the transverse condition $\bar{\nabla}_{\alpha}h^{\alpha}_{~\mu}=0$ gives
\be
\bar{\nabla}_{\alpha}h^{\alpha}_{~y}\propto(y^2-1)f_y''(y)+2yf_y'(y)-\left(2+\frac{n^2}{y^2-1}\right)f_y(y)=0\,,
\ee
from which one can derive the solution
\be\label{fyARN}
f_y(y)=C_1n\left(\frac{y-1}{y+1}\right)^{\frac{|n|}{2}}(|n|+y)\,,~~~~n\in\mathbb{Z}\,,
\ee
with an arbitrary constant $C_1$, that is regular on the horizon and vanishes for $n=0$. Note that in deriving the above solution, we do not resort to the form of the function $f_{\phi}(y)$, which is different form the rotating black hole case analyzed in Sec. \ref{sec2.2}. This is highly due to the near horizon extremal geometry \eqref{g0ARN} is simply a direct product of AdS$_2$ and S$^2$ with warping factors. The function $f_{\phi}(y)$ can be solved independently from
\be
\bar{\nabla}_{\alpha}h^{\alpha}_{~\phi}\propto(y^2-1)f_{\phi}''(y)+2yf_{\phi}'(y)-\frac{n^2}{y^2-1}f_{\phi}(y)=0\,.
\ee
The regular solution takes the form
\be\label{fphiARN}
f_{\phi}(y)=C_2\left(\frac{y-1}{y+1}\right)^{\frac{|n|}{2}}\,,~~~~n\in\mathbb{Z}\,,
\ee
where $C_2$ is an arbitrary constant. So the graviton zero modes are now generated by the following vector fields
\be
\xi=e^{in\tau}\left(\frac{y-1}{y+1}\right)^{\frac{|n|}{2}}\left(C_1\frac{i\left(|n|(|n|+y)+y^2-1\right)}{y^2-1}\frac{\p}{\p\tau}+C_1n(|n|+y)\frac{\p}{\p y}+C_2\frac{\p}{\p\phi}\right)\,,\label{xiARN}
\ee
with integer valued $n$. These vector fields are indeed large diffeomorphisms since they do not die off at the asymptotic boundary $y\to\infty$ and thus can not be gauged away. The existence of the two arbitrary constants $C_1$ and $C_2$ indicates that there are two linearly independent graviton zero modes.

The graviton zero modes which are generated by the $\xi$ with $C_2=0$ and $C_1\neq0$ are call tensor modes. By proper choosing the nonzero constant $C_1$, the explicit forms of the tensor modes are
\begin{align}
h_{\mu\nu}^{t(n)}dx^{\mu}dx^{\nu}&=e^{in\tau}\frac{\sqrt{3|n|(n^2-1)}(1-\mu_0^2)}{2\sqrt{2}\pi\sqrt{3+\mu_0^2}(1-\mu_0\cos\theta)^2(y^2-1)}\left(\frac{y-1}{y+1}\right)^{\frac{|n|}{2}}\nn\\
&\times\left(-(y^2-1)d\tau^2+2i\frac{|n|}{n}d\tau dy+\frac{1}{y^2-1}dy^2\right)\,,~~~~|n|\geq2\,,\label{tzmARN}
\end{align}
and these rank-2 tensors satisfy the following orthonormal condition
\be
\int d^4x\sqrt{\bar{g}}h_{\alpha\beta}^{t(m)*}\bar{g}^{\alpha\mu}\bar{g}^{\beta\nu}h_{\mu\nu}^{t(n)}=\delta^{mn}\,.
\ee
The $n=0, \pm1$ modes are excluded because the corresponding vector fields are Killing vectors on the background and do not produce fluctuations. The graviton zero modes which are generated by the $\xi$ with $C_1=0$ and $C_2\neq0$ are called vector modes, whose explicit forms can be written as
\begin{align}
h_{\mu\nu}^{v(n)}dx^{\mu}dx^{\nu}&=e^{in\tau}\frac{in\mu_0^2\sin^2\theta}{2\sqrt{2}\pi\sqrt{|n|(\mu_0(\mathrm{arctanh}\mu_0)-\mu_0^2)}(y^2-1)}\left(\frac{y-1}{y+1}\right)^{\frac{|n|}{2}}\nn\\
&\times\left((y^2-1)d\tau d\phi-i\frac{|n|}{n}dyd\phi\right)\,,~~~~|n|\geq1\,.\label{vzmARN}
\end{align}
Here the nonzero constant $C_2$ is chosen so that the vector modes are orthonormalized
\be
\int d^4x\sqrt{\bar{g}}h_{\alpha\beta}^{v(m)*}\bar{g}^{\alpha\mu}\bar{g}^{\beta\nu}h_{\mu\nu}^{v(n)}=\delta^{mn}\,.
\ee
Note that for the vector modes, we have $|n|\geq1$. The $n=0$ mode vanishes due to the fact that $\p_{\phi}$ is the rotational Killing vector of the background. One can check that both \eqref{tzmARN} and \eqref{vzmARN} belong to the kernel of the operator $\bar{\Delta}[\bar{g}, \bar{A}]$. It is worth noting that in the absence of acceleration, the Reissner-Nordstr\"om black hole has three sets of vector modes since there are three independent Killing vectors on S$^2$~\cite{Banerjee:2023quv, Blacker:2025zca}. The presence of acceleration deforms the S$^2$ and kills two of the Killing vectors, leaving only the translational symmetry along $\phi$. So there is only one set of the vector modes \eqref{vzmARN} in the accelerating case.

In addition to the graviton zero modes, we also have photon zero modes. The photon zero modes, which satisfy $\left(\bar{P}[\bar{g}, \bar{A}]a\right)_{\mu}=0$, are the residual gauge transformations under the Lorentz gauge. For this reason, we can write the ansatz for the photon zero modes as
\be\label{pzmazARN}
a_{\mu}=\p_{\mu}\Phi(\tau, y)\,,~~~~\Phi(\tau, y)=e^{in\tau}f(y)\,,~~~~n\in\mathbb{Z}\,,
\ee
with a function $f(y)$ to be determined. The Lorentz gauge then requires
\be
\bar{\nabla}_{\mu}a^{\mu}\propto(y^2-1)f''(y)+2yf'(y)-\frac{n^2}{y^2-1}f(y)=0\,.
\ee
The solution to the above equation that is regular on the horizon can be written as
\be\label{fayARN}
f(y)=C_3\left(\frac{y-1}{y+1}\right)^{\frac{|n|}{2}}\,,~~~~n\in\mathbb{Z}\,,
\ee
where $C_3$ is an arbitrary constant. Using \eqref{pzmazARN} and \eqref{fayARN}, we find the photon zero modes
\be
a_{\mu}^{(n)}dx^{\mu}=e^{in\tau}\frac{in\sqrt{1-\mu_0^2}}{2\sqrt{2}\pi \sqrt{|n|}r_0(y^2-1)}\left(\frac{y-1}{y+1}\right)^{\frac{|n|}{2}}\left((y^2-1)d\tau-i\frac{|n|}{n}dy\right)\,,~~~~|n|\geq1\,,\label{pzmARN}
\ee
with orthonormal property
\be
\int d^4x\sqrt{\bar{g}}a_{\mu}^{(m)*}\bar{g}^{\mu\nu}a_{\nu}^{(n)}=\delta^{mn}\,.
\ee
It is straightforward to check that $\eqref{pzmARN}$ is in the kernel of the operator $\bar{P}[\bar{g}, \bar{A}]$. For a torsion free theory, the photon zero modes determined by \eqref{pzmazARN} always give symmetric $\bar{\nabla}_{\mu}a_{\nu}$. It is obvious that the mixing term \eqref{koMARN} vanishes for such zero modes. So the previous derived tensor, vector and photon zero modes indeed belong to the kernel of the matrix operator in \eqref{boARN}.

The path integral over the zero modes on the extremal background still leads to an infrared divergence of the one-loop partition function
\be
Z\propto\int_{\mathrm{tensor~modes}}[Dh^t]\int_{\mathrm{vector~modes}}[Dh^v]\int_{\mathrm{photon~zero~modes}}[Da]\to\infty\,.
\ee
To get finite quantum corrections, we need to perturb the background extremal configuration to the near extremal one, so that the eigenvalues of the zero modes acquire small perturbations and the path integral of the zero modes remains finite.

\subsection{Quantum corrections to the entropy}\label{sec3.3}
The one-loop partition function for the Einstein-Maxwell theory \eqref{ZexpanARN} requires the functional integrations over metric and gauge field fluctuations. If one perform the path integral of the metric fluctuation $h$ first, then the one-loop partition function behaves like
\be
Z\sim\mathrm{det}(\bar{\Delta})^{-\frac{1}{2}}\int[Da]\mathrm{exp}\left[-\int d^4x\sqrt{\bar{g}}a^*\left(\bar{P}-\frac{1}{4}\bar{O}_{\mathrm{int}}^{\dagger}\bar{\Delta}^{-1}\bar{O}_{\mathrm{int}}\right)a\right]\,.
\ee
It is clear that the one-loop corrections to the partition function are determined by the eigenvalues of the operators $\bar{\Delta}$ and $\bar{P}-(\bar{O}_{\mathrm{int}}^{\dagger}\bar{\Delta}^{-1}\bar{O}_{\mathrm{int}})/4$. For the zero modes \eqref{tzmARN}, \eqref{vzmARN} and \eqref{pzmARN}, their corresponding eigenvalues vanish and the one-loop partition function diverges.

To regularize the divergence, one should take into account the $\mathcal{O}(T)$ corrections \eqref{g1ARN} and \eqref{A1ARN} to the extremal metric \eqref{g0ARN} and the extremal gauge field \eqref{A0ARN}, respectively
\be
\bar{g}\to\bar{g}+\delta g\,,~~~~\bar{A}\to\bar{A}+\delta A\,.
\ee
This will perturb the operators $\bar{\Delta}$ and $\bar{P}-(\bar{O}_{\mathrm{int}}^{\dagger}\bar{\Delta}^{-1}\bar{O}_{\mathrm{int}})/4$ and correct the the eigenvalues of the zero modes to nonzero ones. According to the perturbation theory at first order, the eigenvalue correction can be evaluated as the expectation value of the perturbed operator with respect to the original orthonormal eigenstates. However, from \eqref{koMARN}, it is obvious that the photon zero modes \eqref{pzmARN} satisfy
\be
\bar{O}_{\mathrm{int}}^{\alpha\beta\mu}[\bar{g}, \bar{A}]a_{\mu}=0\,.
\ee
So the variation of the operator $(\bar{O}_{\mathrm{int}}^{\dagger}\bar{\Delta}^{-1}\bar{O}_{\mathrm{int}})/4$ will not contribute to the first-order eigenvalue correction. Thus, to figure out the eigenvalue corrections, we only need to consider the variations of the operators $\bar{\Delta}$ and $\bar{P}$. Given the orthonormalized zero modes \eqref{tzmARN}, \eqref{vzmARN} and \eqref{pzmARN}, the corrections to their eigenvalues can be formulated as
\begin{align}
\delta\Lambda_n^t&=\int d^4x\sqrt{\bar{g}}h_{\alpha\beta}^{t(n)*}\delta\Delta^{\alpha\beta,\mu\nu}h_{\mu\nu}^{t(n)}\,,\label{dlt}\\
\delta\Lambda_n^v&=\int d^4x\sqrt{\bar{g}}h_{\alpha\beta}^{v(n)*}\delta\Delta^{\alpha\beta,\mu\nu}h_{\mu\nu}^{v(n)}\,,\label{dlv}\\
\delta\lambda_n&=\int d^4x\sqrt{\bar{g}}a_{\mu}^{(n)*}\delta P^{\mu\nu}a_{\nu}^{(n)}\,,\label{dla}
\end{align}
where we denote $\Lambda_n^t$, $\Lambda_n^v$ and $\lambda_n$ as the eigenvalues of the tensor modes $h^{t(n)}$, vector modes $h^{v(n)}$ and photon zero modes $a^{(n)}$, respectively. The above integrations are straightforward. After some algebra, we find
\begin{align}
\delta\Lambda_n^t&=\frac{3|n|(1-\mu_0^2)\sqrt{1+\mu_0^2}}{16r_0(3+\mu_0^2)}T+\mathcal{O}(T^2)\,,~~~|n|\geq2\,,\\
\delta\Lambda_n^v&=\frac{|n|\mu_0^3\sqrt{1+\mu_0^2}}{48r_0(1-\mu_0^2)(\mathrm{arctanh}\mu_0-\mu_0)}T+\mathcal{O}(T^2)\,,~~~|n|\geq1\,,\\
\delta\lambda_n&=\frac{|n|\left(\mu_0-(1-\mu_0^2)\mathrm{arctanh}\mu_0\right)\sqrt{1+\mu_0^2}}{8r_0\mu_0}T+\mathcal{O}(T^2)\,,~~~|n|\geq1\,.
\end{align}

By summing up the contributions from the three classes of zero modes in the path integral and using the zeta function regularizations
\be
\prod_{n\geq2}\frac{\rho}{n}=\frac{1}{\rho^{\frac{3}{2}}\sqrt{2\pi}}\,,~~~~\prod_{n\geq1}\frac{\sigma}{n}=\frac{1}{\sigma^{\frac{1}{2}}\sqrt{2\pi}}\,,
\ee
for real $\rho$ and $\sigma$, we get the leading quantum correction to the near extremal entropy
\begin{align}
\delta\log Z&=-\frac{1}{2}\sum_{|n|\geq2}\log\left(\delta\Lambda_n^t\right)-\frac{1}{2}\sum_{|n|\geq1}\log\left(\delta\Lambda_n^v\right)-\frac{1}{2}\sum_{|n|\geq1}\log\left(\delta\lambda_n\right)\\
&=-\log\prod_{n\geq2}\left(\delta\Lambda_n^t\right)-\log\prod_{n\geq1}\left(\delta\Lambda_n^v\right)-\log\prod_{n\geq1}\left(\delta\lambda_n\right)\\
&=\frac{3}{2}\log\frac{T}{T_q^t}+\frac{1}{2}\log\frac{T}{T_q^v}+\frac{1}{2}\log\frac{T}{T_q^p}+\log\left(\frac{3}{2048}\left(\frac{1}{\pi}\right)^{\frac{3}{2}}\right)+\cdots\,,
\end{align}
where
\be
T_q^t=\frac{r_0(3+\mu_0^2)}{(1-\mu_0^2)\sqrt{1+\mu_0^2}}\,,~~T_q^v=\frac{r_0(1-\mu_0^2)(\mathrm{arctanh}\mu_0-\mu_0)}{\mu_0^3\sqrt{1+\mu_0^2}}\,,~~T_q^p=\frac{r_0\mu_0}{\left(\mu_0-(1-\mu_0^2)\mathrm{arctanh}\mu_0\right)\sqrt{1+\mu_0^2}}\,.
\ee
Hence, the near extremal partition function of the accelerating Reissner-Nordstr\"om black hole is corrected as
\be
Z(T)\approx T^{\frac{5}{2}}e^{S_0}+\mathrm{higher~order~terms}\,,\label{ZTARN}
\ee
where $S_0$ is the extremal entropy. Note that the temperature dependence in the leading term of \eqref{ZTARN} is different from that in the accelerating Kerr case \eqref{ZT}. This is because the absence of rotation and the presence of the gauge field each leads to additional zero modes and contributes $T^{1/2}$ factor in the partition function. Accordingly, the black hole thermal energy above extremality in this case is dominated by a term scaling as $(5/2)T$ at low temperature which replace it on the same footing as the Hawking radiation.

\section{Accelerating Kerr-Newmann black hole}\label{sec4}
We finally analyze the low-temperature quantum corrections to the thermodynamics of the C-metric with both rotation and charge. As we will see, by turning on the rotation, the gravitational zero modes of vector type for the accelerating Reissner-Nordstr\"om black hole are no longer the case. The spacetime line element and the $U(1)$ gauge field for the accelerating Kerr-Newmann black hole in the Boyer-Lindquist coordinate system take the following forms~\cite{Hong:2004dm, Griffiths:2005se, Griffiths:2005qp}
\begin{align}
ds^2&=-\frac{\Delta_r}{H^2\rho^2}\left(\frac{dt}{b}-a\sin^2\theta d\phi\right)^2+\frac{\rho^2}{H^2}\left(\frac{dr^2}{\Delta_r}+\frac{d\theta^2}{\Delta_{\theta}}\right)+\frac{\Delta_{\theta}a^2\sin^2\theta}{H^2\rho^2}\left(\frac{dt}{b}-\frac{r^2+a^2}{a}d\phi\right)^2\,,\label{BLgAKN}\\
A&=-\frac{qr}{\rho^2}\left(\frac{dt}{b}-a\sin^2\theta d\phi\right)+\Phi dt\,,\label{AAKN}
\end{align}
where
\begin{align}
H&=1-\alpha r\cos\theta\,,~~\rho^2=r^2+a^2\cos^2\theta\,,\nn\\
\Delta_{\theta}&=1-2\alpha m\cos\theta+(\alpha^2a^2+\alpha^2q^2)\cos^2\theta\,,\\
\Delta_r&=(1-\alpha^2r^2)X\,,~~X=r^2-2mr+a^2+q^2\,.\nn
\end{align}
$m$, $a$, $q$, and $\alpha$ are the black hole mass, angular momentum, charge and acceleration parameters, respectively. $b$ is the time scaling parameter and $\Phi$ is the charge potential at the horizon. The outer and inner horizon radius $r_+$ and $r_-$ are now given by
\be\label{rpmAKN}
r_+=m+\sqrt{m^2-a^2-q^2}\,,~~~~r_-=m-\sqrt{m^2-a^2-q^2}\,.
\ee
The relevant thermodynamic quantities of the accelerating Kerr-Newmann black hole are listed as the following~\cite{Anabalon:2018qfv}
\begin{align}
M&=\frac{m(1-\alpha^2a^2)}{b(1+\alpha^2a^2)}\,,~~~~T=\frac{(r_+-r_-)(1-\alpha^2r_+^2)}{4\pi b(r_+^2+a^2)}\,,~~~~S=\frac{\pi(r_+^2+a^2)}{1-\alpha^2r_+^2}\,,\nn\\
\Omega&=\frac{a}{b(r_+^2+a^2)}-\frac{\alpha^2a}{b(1+\alpha^2a^2)}\,,~~~~J=ma\,,~~~~\Phi=\frac{qr_+}{b(r_+^2+a^2)}\,,~~~~Q=q\,,\label{thermqAKN}\\
\lambda_{\pm}&=\frac{r_+}{b(1\mp\alpha r_+)}-\frac{M}{1+\alpha^2a^2+\alpha^2q^2}\pm\frac{\alpha a^2}{b(1+\alpha^2a^2)}\,,~~~~\mu_{\pm}=\frac{1}{4}(\pm2\alpha m-\alpha^2a^2-\alpha^2q^2)\,.\nn
\end{align}
Here the notations of the symbols above follow their earlier definitions. The net string tension $\mu$ satisfies an equation analogous to Newton's second law
\be\label{munetAKN}
\mu=\mu_+-\mu_-=m\alpha\,.
\ee
The thermodynamic first law holds for the above quantities, i.e.,
\be
dM=TdS+\Omega dJ+\Phi dQ-\lambda_+d\mu_+-\lambda_-d\mu_-\,,
\ee
given that
\be
b=\frac{\sqrt{(1-\alpha^2a^2)(1+\alpha^2a^2+\alpha^2q^2)}}{1+\alpha^2a^2}\,.
\ee

\subsection{Near extremal thermodynamics and geometry}\label{sec4.1}
The near extremal expansions are now performed by taking the temperature $T$ as small parameter while keeping the angular momentum, charge and net string tension at their extremal values
\be
J_0=r_0\sqrt{r_0^2-q_0^2}\,,~~~~Q_0=q_0\,,~~~~\mu_0=r_0\alpha\,,
\ee
where $r_0$ is the extremal horizon radius and $q_0$ denotes the extremal charge. In this canonical ensemble, the horizon radius, black hole mass and entropy have the following small temperature expansions
\begin{align}
r_+&=r_0+\frac{2\pi r_0(2r_0^2-q_0^2)\sqrt{q_0^2\mu_0^2(1+\mu_0^2)+r_0^2(1-\mu_0^4)}}{\left(r_0^2+(r_0^2-q_0^2)\mu_0^2\right)(1-\mu_0^2)}T+\mathcal{O}(T^2)\,,\label{rpexpanAKN}\\
r_-&=r_0-\frac{2\pi r_0(2r_0^2-q_0^2)\sqrt{q_0^2\mu_0^2(1+\mu_0^2)+r_0^2(1-\mu_0^4)}}{\left(r_0^2+(r_0^2-q_0^2)\mu_0^2\right)(1-\mu_0^2)}T+\mathcal{O}(T^2)\,,\label{rmexpanAKN}\\
M&=\sqrt{\frac{r_0^2-(r_0^2-q_0^2)\mu_0^2}{1+\mu_0^2}}+\frac{2\pi^2(q_0^2-2r_0^2)\left(r_0^6+r_0^6\mu_0^2-r_0^2(2q_0^4-5q_0^2r_0^2+3r_0^4)\mu_0^4+(r_0^2-q_0^2)^3\mu_0^6\right)}{\left(\left(r_0^2+(r_0^2-q_0^2)\mu_0^2\right)(1-\mu_0^2)\right)^2\sqrt{\left(r_0^2-(r_0^2-q_0^2)\mu_0^2\right)(1+\mu_0^2)}}T^2+\mathcal{O}(T^3)\,,\label{MexpanAKN}\\
S&=\frac{\pi(2r_0^2-q_0^2)}{1-\mu_0^2}+\frac{4\pi^2(2r_0^2-q_0^2)\sqrt{\left(r_0^2-(r_0^2-q_0^2)\mu_0^2\right)(1+\mu_0^2)}}{(1-\mu_0^2)^3}T+\mathcal{O}(T^2)\,.\label{SexpanAKN}
\end{align}
As a consistency check, the chargeless limit $q_0\to0$ of the above expansions recover the accelerating Kerr black hole case \eqref{rpexpan}, \eqref{rmexpan}, \eqref{Mexpan}, and \eqref{Sexpan} while the static limit $q_0\to r_0$ reproduce the accelerating Reissner-Nordstr\"om black hole case \eqref{rpexpanARN}, \eqref{rmexpanARN}, \eqref{MexpanARN}, and \eqref{SexpanARN}.

To compute the low-temperature quantum corrections to the thermodynamic quantities, we need to figure out the near extremal configuration which can be obtained by the following coordinate transformation
\begin{align}
r&=r_++\frac{2\pi r_0(2r_0^2-q_0^2)\sqrt{q_0^2\mu_0^2(1+\mu_0^2)+r_0^2(1-\mu_0^4)}}{\left(r_0^2+(r_0^2-q_0^2)\mu_0^2\right)(1-\mu_0^2)}T(y-1)\,,\label{nexct1AKN}\\
t&=-i\frac{b}{r_0}\frac{r_0^2+(r_0^2-q_0^2)\mu_0^2}{\sqrt{\left(r_0^2-(r_0^2-q_0^2)\mu_0^2\right)(1+\mu_0^2)}}\frac{\tau}{2\pi T}\,,\label{nexct2AKN}\\
\phi&=\frac{2r_0\sqrt{r_0^2-q_0^2}}{(2r_0^2-q_0^2)(1-\mu_0^2)}\varphi-i\frac{\sqrt{r_0^2-q_0^2}}{2r_0^2-q_0^2}\frac{r_0^2+(r_0^2-q_0^2)\mu_0^2}{\sqrt{\left(r_0^2-(r_0^2-q_0^2)\mu_0^2\right)(1+\mu_0^2)}}\frac{\tau}{2\pi r_0T}+i\frac{2r_0\sqrt{r_0^2-q_0^2}}{(2r_0^2-q_0^2)(1-\mu_0^2)}\tau\,,\label{nexct3AKN}
\end{align}
where $\tau$ is the Euclidean time with period $2\pi$, $y\geq1$ and $\varphi$ is identified as
\be
\varphi\sim\varphi+\frac{\pi(2r_0^2-q_0^2)(1-\mu_0^2)}{r_0\sqrt{r_0^2-q_0^2}}\,.
\ee
When substituting the transformations \eqref{nexct1AKN}, \eqref{nexct2AKN} and \eqref{nexct3AKN} into the metric \eqref{BLgAKN} and the gauge field \eqref{AAKN} and taking the small-temperature limit with fixed angular momentum, charge and net string tension at their extremal values, we find the leading $\mathcal{O}(T^0)$ terms
\begin{align}
d\bar{s}^2=&\frac{r_0^2+(r_0^2-q_0^2)\cos^2\theta}{(1-\mu_0^2)(1-\mu_0\cos\theta)^2}\left((y^2-1)d\tau^2+\frac{1}{y^2-1}dy^2\right)+\frac{r_0^2+(r_0^2-q_0^2)\cos^2\theta}{(1-\mu_0\cos\theta)^4}d\theta^2\nn\\
&+\frac{4r_0^2(r_0^2-q_0^2)\sin^2\theta}{(1-\mu_0^2)\left(r_0^2+(r_0^2-q_0^2)\cos^2\theta\right)}\left(d\varphi-i(y-1)d\tau\right)^2\,,\label{g0AKN}\\
\bar{A}=&\frac{q_0\left(r_0^2-(r_0^2-q_0^2)\cos^2\theta\right)}{(1-\mu_0^2)\left(r_0^2+(r_0^2-q_0^2)\cos^2\theta\right)}\left(\frac{2r_0^2(r_0^2-q_0^2)\sin^2\theta}{(2r_0^2-q_0^2)\left(r_0^2-(r_0^2-q_0^2)\cos^2\theta\right)}d\varphi-i(y-1)d\tau\right)\,.\label{A0AKN}
\end{align}
The subleading $\mathcal{O}(T)$ perturbation of the metric is
\begin{align}
\frac{\delta(ds^2)}{T}=&\frac{4\pi(2r_0^2-q_0^2)\mu_0\sqrt{\left(r_0^2-(r_0^2-q_0^2)\mu_0^2\right)(1+\mu_0^2)}y\cos\theta}{\left(r_0^2+(r_0^2-q_0^2)\mu_0^2\right)(1-\mu_0^2)(1-\mu_0\cos\theta)}d\bar{s}^2\nn\\
&+g_1(y, \theta)d\tau^2+g_2(y, \theta)dy^2+g_3(y, \theta)d\theta^2+\left(g_4(y, \theta)d\varphi+g_5(y, \theta)d\tau\right)\left(d\varphi-i(y-1)d\tau\right)\,,\label{g1AKN}
\end{align}
where
\begin{align}
g_1(y, \theta)&=\frac{4\pi\sqrt{\left(r_0^2-(r_0^2-q_0^2)\mu_0^2\right)(1+\mu_0^2)}(y-1)}{\left(r_0^2+(r_0^2-q_0^2)\mu_0^2\right)(1-\mu_0^2)^3(1-\mu_0\cos\theta)^2}\Big[-\left(2(r_0^2-q_0^2)r_0^2\mu_0^2+r_0^2q_0^2\right)(y^2+y)\nn\\
&+(r_0^2-q_0^2)(2r_0^2-q_0^2\mu_0^2)(y^2+y)\cos^2\theta+2\left(r_0^2(1+\mu_0^2)-q_0^2\mu_0^2\right)\left(r_0^2+(r_0^2-q_0^2)\cos^2\theta\right)\Big]\,,\\
g_2(y, \theta)&=\frac{4\pi\sqrt{\left(r_0^2-(r_0^2-q_0^2)\mu_0^2\right)(1+\mu_0^2)}}{\left(r_0^2+(r_0^2-q_0^2)\mu_0^2\right)(1-\mu_0^2)^3(y^2-1)(y+1)(1-\mu_0\cos\theta)^2}\Big[r_0^2(2r_0^2-q_0^2)(y^2+y)\nn\\
&+(2r_0^2-q_0^2)(r_0^2-q_0^2)\mu_0^2(y^2+y)\cos^2\theta-2\left(r_0^2(1+\mu_0^2)-q_0^2\mu_0^2\right)\left(r_0^2+(r_0^2-q_0^2)\cos^2\theta\right)\Big]\,,\\
g_3(y, \theta)&=\frac{4\pi r_0^2(2r_0^2-q_0^2)\sqrt{\left(r_0^2-(r_0^2-q_0^2)\mu_0^2\right)(1+\mu_0^2)}y}{\left(r_0^2+(r_0^2-q_0^2)\mu_0^2\right)(1-\mu_0^2)(1-\mu_0\cos\theta)^4}\,,\\
g_4(y, \theta)&=\frac{8\pi r_0^4(r_0^2-q_0^2)\sqrt{\left(r_0^2-(r_0^2-q_0^2)\mu_0^2\right)(1+\mu_0^2)}y\left((r_0^2-q_0^2)\sin^2(2\theta)+2\sin^2\theta\right)}{\left(r_0^2+(r_0^2-q_0^2)\mu_0^2\right)(1-\mu_0^2)^3\left(r_0^2+(r_0^2-q_0^2)\cos^2\theta\right)^2}\,,\\
g_5(y, \theta)&=\frac{4\pi ir_0^2(r_0^2-q_0^2)\sqrt{\left(r_0^2-(r_0^2-q_0^2)\mu_0^2\right)(1+\mu_0^2)}\sin^2\theta}{\left(r_0^2+(r_0^2-q_0^2)\mu_0^2\right)(1-\mu_0^2)^4\left(r_0^2+(r_0^2-q_0^2)\cos^2\theta\right)^2(1-\mu_0\cos\theta)^2}\nn\\
&\times\Big[-2(1-\mu_0^2)^2(y^2-1)\left(r_0^2+(r_0^2-q_0^2)\cos^2\theta\right)^2+r_0q_0\left(-1+9\mu_0-2(1-\mu_0^2)(y^2-2y)\right)(1-\mu_0\cos\theta)^2\nn\\
&-\left(2r_0^4\left(5+3\mu_0^2-2(1-\mu_0^2)y^2\right)+q_0^2(r_0^2-q_0^2)\left(1-9\mu_0^2-2(1-\mu_0^2)y^2\right)\cos^2\theta\right)(1-\mu_0\cos\theta)^2\nn\\
&-2r_0^2(r_0^2-q_0^2)\left(5+3\mu_0+2(1-\mu_0^2)(y^2-2y)\right)(1-\mu_0\cos\theta)^2\cos^2\theta\Big]\,.
\end{align}
and the $\mathcal{O}(T)$ perturbation of the gauge field is
\begin{align}
\frac{\delta(A)}{T}&=a_1(y, \theta)d\tau+a_2(y, \theta)d\varphi\,,\label{A1AKN}
\end{align}
where
\begin{align}
a_1(y, \theta)&=\frac{-\pi ir_0^2q_0\sqrt{\left(r_0^2(1-\mu_0^2)+q_0^2\mu_0^2\right)(1+\mu_0^2)}}{(2r_0^2-q_0^2)\left(r_0^2+(r_0^2-q_0^2)\mu_0^2\right)(1-\mu_0^2)^3\left(r_0^2+(r_0^2-q_0^2)\cos^2\theta\right)^2}\nn\\
&\times\Big[2r_0^6\left((1-\mu_0^2)(-4y^2+4y)+5\mu_0^2+3\right)+r_0^4q_0^2\left((1-\mu_0^2)(8y^2-12y)-21\mu_0^2-3\right)\nn\\
&+r_0^2q_0^4\left((1-\mu_0^2)(-2y^2+4y)+9\mu_0^2-1\right)-2r_0^2q_0^2(r_0^2-q_0^2)\left((1-\mu_0^2)(12y^2-8y)+9\mu_0^2-1\right)\cos^2\theta\nn\\
&-8r_0^4(r_0^2-q_0^2)(1-\mu_0^2)(-3y^2+2y+1)\cos^2\theta-q_0^4(r_0^2-q_0^2)\left((1-\mu_0^2)(-6y^2+4y)-9\mu_0^2+1\right)\cos^2\theta\nn\\
&-2r_0^2(r_0^2-q_0^2)^2\left(-4(1-\mu_0^2)y+\mu_0^2+7\right)\cos^4\theta+q_0^2(r_0^2-q_0^2)^2\left(-4(1-\mu_0^2)y+3\mu_0^2+5\right)\cos^4\theta\Big]\,,\\
a_2(y, \theta)&=-\frac{4\pi r_0^2q_0(r_0^2-q_0^2)\sqrt{r_0^2(1-\mu_0^4)+q_0^2\mu_0^2(1+\mu_0^2)}y\left(r_0^2-(r_0^2-q_0^2)\cos^2\theta\right)\sin^2\theta}{\left(r_0^2(1+\mu_0^2)-q_0^2\mu_0^2\right)(1-\mu_0^2)^2\left(r_0^2+(r_0^2-q_0^2)\cos^2\theta\right)^2}\,.
\end{align}

\subsection{Euclidean path integral near extremality and zero modes}\label{sec4.2}
The one-loop partition function of the accelerating Kerr-Newmann black hole shares the same form as \eqref{ZexpanARN} with path integral over metric and gauge field fluctuations. The saddle-point here is the extremal configurations given by \eqref{g0AKN} and \eqref{A0AKN}.

For the same reason as discussed in the previous sections, among all the possible fluctuations, zero modes contribute leading quantum corrections to the partition function at low temperature. The zero modes are divided into graviton zero modes satisfying $\left(\bar{\Delta}[\bar{g}, \bar{A}]h\right)_{\mu\nu}=0$ and photon zero modes satisfying $\left(\bar{P}[\bar{g}, \bar{A}]a\right)_{\mu}=0$. The graviton zero modes here are generated by the same vector fields given by \eqref{xi}. This can be easily checked by setting the same ansatz \eqref{xiansatz} for the large diffeomorphisms and successively consider the following constraints of the zero modes
\begin{align}
&h^{\alpha}_{~\alpha}=0\,,\\
&\frac{(2r_0^2-q_0^2)(1-\mu_0^2)}{\left(r_0^2(1+\mu_0)-q_0^2\right)(1+\mu_0)}\p_{\theta}\left(\bar{\Delta}[\bar{g}, \bar{A}]h\right)_{y\theta}|_{\theta=0}\nn\\
&+\frac{r_0^2}{r_0^2(1+\mu_0^2)-q_0^2}\p_{\theta}\left(\bar{\Delta}[\bar{g}, \bar{A}]h\right)_{y\theta}|_{\theta=\frac{\pi}{2}}=0\,,\\
&\bar{\nabla}_{\alpha}h^{\alpha}_{~y}=0\,,
\end{align}
By taking the Lie derivative of the metric \eqref{g0AKN} along \eqref{xi}, one finds the following orthonormalized graviton zero modes
\begin{align}
h_{\mu\nu}^{(n)}dx^{\mu}dx^{\nu}&=e^{in\tau}\frac{\sqrt{3|n|(n^2-1)}(1-\mu_0^2)\left(r_0^2+(r_0^2-q_0^2)\cos^2\theta\right)\left(\frac{y-1}{y+1}\right)^{\frac{|n|}{2}}}{2\sqrt{2}\pi\sqrt{(2r_0^2-q_0^2)\left(4r_0^2(1+\mu_0^2)-q_0^2(1+3\mu_0^2)\right)}(1-\mu_0\cos\theta)^2(y^2-1)}\nn\\
&\times\left(-(y^2-1)d\tau^2+2i\frac{|n|}{n}d\tau dy+\frac{1}{y^2-1}dy^2\right)\,,~~~~|n|\geq2\,.\label{zeromodesAKN}
\end{align}
In the chargeless limit $q_0\to0$, the above metric fluctuations coincide with the zero modes \eqref{zeromodes} in the accelerating Kerr case. The static limit $q_0\to r_0$ of the above metric fluctuations only reproduce the tensor modes \eqref{tzmARN} in the accelerating Reissner-Nordstr\"om case. We do not find vector modes here in the rotating case. This is highly related to the near horizon geometries of the extremal black hole. When turning on the rotation, the near horizon geometry of the extremal black hole becomes a warped and twisted product of AdS$_2$ and S$^2$ rather than a direct product, and the $\varphi$ components of the large diffeomorphisms are no longer independent of their $\tau$ and $y$ components. So there is no independent vector mode in the rotating case.

In addition to the graviton zero modes, we also need to consider the contribution from the photon zero modes due to the presence of the gauge field \eqref{AAKN}. Since we are using the Lorentz gauge, the determination of the photon zero modes follows the same procedure in Sec. \ref{sec3.2}. For the background extremal geometry \eqref{g0AKN}, the orthonormalized photon zero modes can be written as
\be
a_{\mu}^{(n)}dx^{\mu}=e^{in\tau}\frac{in\sqrt{1-\mu_0^2}\left(\frac{y-1}{y+1}\right)^{\frac{|n|}{2}}}{2\sqrt{2}\pi \sqrt{|n|}\sqrt{2r_0^2-q_0^2}(y^2-1)}\left((y^2-1)d\tau-i\frac{|n|}{n}dy\right)\,,~~~~|n|\geq1\,.\label{pzmAKN}
\ee
It is clear that the static limit $q_0\to r_0$ of the above gauge fluctuations reproduce the photon zero modes \eqref{pzmARN} in the accelerating Reissner-Nordstr\"om case. The mixing term \eqref{koMARN} also vanishes here, so the fluctuations \eqref{zeromodesAKN} and \eqref{pzmAKN} are zero modes of the matrix operator defined in \eqref{boARN}.

\subsection{Quantum corrections to the entropy}\label{sec4.3}
Finite quantum corrections to the partition function are obtained by adding the finite temperature corrections \eqref{g1AKN} and \eqref{A1AKN} to their extremal configurations \eqref{g0AKN} and \eqref{A0AKN} in the path integral over the zero modes \eqref{zeromodesAKN} and \eqref{pzmAKN}. The corrections to the eigenvalues of the graviton and photon zero modes are given by the formulas \eqref{dlt} and \eqref{dla}, respectively. After substituting the zero modes and the near extremal configurations into \eqref{dlt} and \eqref{dla} and retaining all the $\mathcal{O}(T)$ terms, the integrations yield
\begin{align}
\delta\Lambda_n=&\frac{3|n|r_0(1-\mu_0^2)\sqrt{r_0^2(1-\mu_0^4)+q_0^2\mu_0^2(1+\mu_0^2)}}{16\left(4r_0^2(1+\mu_0^2)-q_0^2(1+3\mu_0^2)\right)}T+\mathcal{O}(T^2)\,,~~~~|n|\geq2\,,\\
\delta\lambda_n=&\frac{|n|(2r_0^2-q_0^2)\mu_0\sqrt{\left(r_0^2-(r_0^2-q_0^2)\mu_0^2\right)(1+\mu_0^2)}}{8r_0\left(r_0^2+(r_0^2-q_0^2)\mu_0^2\right)\left(r_0^2(1+\mu_0^2)-q_0^2\right)\sqrt{r_0^2-q_0^2}}\nn\\
&\times\left[(2r_0^2-q_0^2)\mathrm{arccot}\left(\frac{r_0}{\sqrt{r_0^2-q_0^2}}\right)\mu_0-r_0\sqrt{r_0^2-q_0^2}(1-\mu_0^2)\mathrm{arctanh}\mu_0\right]T+\mathcal{O}(T^2)\,,~~~~|n|\geq1\,,
\end{align}
where $\delta\Lambda$ and $\delta\lambda_n$ are the corrections to the eigenvalues of the graviton zero modes $h^{(n)}$ and the photon zero modes $a^{(n)}$, respectively. The leading quantum correction to the near extremal entropy can be obtained by summing up the contributions from the two classes of zero modes with the zeta function regularizations
\begin{align}
\delta\log Z&=-\frac{1}{2}\sum_{|n|\geq2}\log\left(\delta\Lambda_n\right)-\frac{1}{2}\sum_{|n|\geq1}\log\left(\delta\lambda_n\right)\\
&=-\log\prod_{n\geq2}\left(\delta\Lambda_n\right)-\log\prod_{n\geq1}\left(\delta\lambda_n\right)\\
&=\frac{3}{2}\log\frac{T}{T_q^g}+\frac{1}{2}\log\frac{T}{T_q^p}+\log\left(\frac{3}{256\pi}\sqrt{\frac{3}{2}}\right)+\cdots\,,
\end{align}
where
\begin{align}
T_q^g=&\frac{4r_0^2(1+\mu_0^2)-q_0^2(1+3\mu_0^2}{r_0(1-\mu_0^2)\sqrt{r_0^2(1-\mu_0^4)+q_0^2\mu_0^2(1+\mu_0^2)}}\,,\\
T_q^p=&\frac{r_0\left(r_0^2+(r_0^2-q_0^2)\mu_0^2\right)\left(r_0^2(1+\mu_0^2)-q_0^2\right)\sqrt{r_0^2-q_0^2}}{(2r_0^2-q_0^2)\mu_0\sqrt{\left(r_0^2-(r_0^2-q_0^2)\mu_0^2\right)(1+\mu_0^2)}}\nn\\
&\times\left[(2r_0^2-q_0^2)\mathrm{arccot}\left(\frac{r_0}{\sqrt{r_0^2-q_0^2}}\right)\mu_0-r_0\sqrt{r_0^2-q_0^2}(1-\mu_0^2)\mathrm{arctanh}\mu_0\right]^{-1}\,.
\end{align}
Accordingly, the thermal partition function of the accelerating Kerr-Newmann black hole acquires low-temperature quantum corrections of the form
\be
Z(T)\approx T^2e^{S_0}+\mathrm{higher~order~terms}\,,\label{ZTAKN}
\ee
where $S_0$ is the extremal entropy. The power of $T$ in the leading term of \eqref{ZTAKN} lies in between the neutral rotating \eqref{ZT} and the charged static \eqref{ZTARN} cases due to absence of the gravitational vector modes.

\section{Summary and discussion}
In this paper, we systematically study the low-temperature quantum corrections to the thermodynamics of the near extremal rotating and charged black holes with acceleration in four dimensions. The quantum corrections originate from the zero modes, which are the residual gauge fluctuations of the fields under gauge fixing, in the Euclidean path integral. Depending on the near horizon extremal geometry and the field contents, the zero modes have different behaviors. The metric fluctuations under large diffeomorphisms are known as the gravitational zero modes and the pure gauge fluctuations of the gauge field are specified as the photon zero modes. Each class of the zero modes is found to contribute a universal logarithmic term to the black hole thermal entropy at low temperature.

We first consider the accelerating Kerr black hole. As discussed in Sec. \ref{sec2}, the thermodynamic quantities defined by using semiclassical analysis exhibit problematic low-temperature behaviors, since the black hole has no adequate thermal energy to radiate even a single Hawking quantum once its temperature drops to sufficiently small values. The way to resolve the inconsistency between the thermal energy and the Hawking radiation at low temperature is to incorporate quantum corrections to the black hole thermal partition function. Among all the fluctuations in the Euclidean path integral of the partition function, zero modes are found to contribute leading quantum corrections at low temperature. Invoking the harmonic gauge, the zero modes in the accelerating Kerr black hole case are those metric fluctuations on the extremal background that satisfy the transverse and traceless conditions. At low temperature, the Euclidean path integral of these zero modes produce a $(3/2)\log T$ term correction to the black hole entropy and a $(3/2)T$ term correction to the black hole mass. The critical temperature below which these quantum corrections become important is found to depend on the extremal horizon radius as well as the acceleration of the black hole. For fixed horizon radius, the critical temperature gets higher as the acceleration of the black hole increases.

We next consider the accelerating Reissner-Nordstr\"om black hole case. As presented in Sec. \ref{sec3}, the small temperature expansion of the black hole mass from semiclassical analysis still points to the mass gap puzzle. Compared to the neutral rotating case, the zero modes structure becomes richer in this charged static case. The near horizon extremal geometry is a direct product of AdS$_2$ and S$^2$ with warping factors. This structure consequently enables us to divide the large diffeomorphisms into two linearly independent classes: one consists of the diffeomorphisms on the AdS$_2$ which generate the gravitational zero modes of tensor type, and the other includes the deformations of the S$^2$ along the AdS$_2$ which correspond to the gravitational zero modes of vector type. It is worth noting that in the absence of acceleration, the Reissner-Nordstr\"om black hole has three sets of vector modes since there are three independent Killing vectors on the sphere. The presence of acceleration kills two of them with only translation along the polar angle remaining. So there is only one set of the vector modes in the accelerating case. At low temperature, the path integral of the tensor modes produce a $(3/2)\log T$ term correction to the black hole entropy while the vector modes contribute a $(1/2)\log T$ term correction to the entropy. In addition, we also have the photon zero modes from the gauge field, which are the large gauge transformations under the Lorentz gauge fixing. The path integral of the photon zero modes gives another $(1/2)\log T$ term correction to the black hole entropy. The total contribution from the zero modes to the black hole entropy in the accelerating Reissner-Nordstr\"om case at low temperature is therefore $(5/2)\log T$.

We finally discuss the low-temperature quantum corrections to the accelerating Kerr-Newmann black hole in Sec. \ref{sec4}. Its near horizon extremal geometry shares a similar structure to the accelerating Kerr case, which is a warped and twisted product of AdS$_2$ and S$^2$. This determines that the gravitational zero modes are purely tensor type. The path integral of the gravitational tensor modes therefore contributes a $(3/2)\log T$ term correction to the black hole entropy at low temperature. Since this is a charged black hole, the gauge field contains also the photon zero modes, whose path integral leads to a $(1/2)\log T$ term correction in the low-temperature expansion of the entropy. So the total contribution from the zero modes to the black hole entropy in the accelerating Kerr-Newmann case at low temperature is $2\log T$.

It seems that the $\log T$ corrections to the entropy of near extremal black holes in Einstein gravity are universal with coefficients depending on the class of the zero modes. For each gravitational zero modes of tensor type, the integer $n$ which labels the infinite tower of the zero modes satisfies $|n|\geq2$, and the coefficient of the corresponding logarithmic correction term is $3/2$. For each gravitational zero modes of vector type, the integer $n$ satisfies $|n|\geq1$, and the coefficient of the corresponding logarithmic term is $1/2$. For the zero modes from gauge field, the infinite tower is also labeled by integers $|n|\geq1$, hence the corresponding coefficient is also $1/2$. For rotating black holes there is no gravitational vector modes.

To sum up, we worked out the universal $\log T$ corrections to the low-temperature thermodynamics of accelerating black holes. The results are similar to those in the nonaccelerating cases like Kerr and Reissner-Nordstr\"om black holes. The acceleration affects the near horizon extremal geometry with warping factors and therefore changes the number of independent Killing vectors on the deformed sphere. This in turn changes the number of sets of independent gravitation vector modes which are induced by the large diffeomorphisms on the deformed sphere.

\section*{Acknowledgment}
We are grateful to Jianxin Lu, Jun Nian and Jingchao Zhang for helpful discussions. This work is supported by the National Natural Science Foundation of China under Grants No. 12105045 and No. 12247103.

\end{document}